\documentclass[10pt]{article}

\usepackage{epsfig}
\usepackage{amssymb}
\usepackage{amsmath}

\renewcommand{\baselinestretch}{2.0}

\begin{document}

\pagestyle{plain}
\pagenumbering{arabic}
\setcounter{page}{1}


\begin{center}

{\Large \bf A Logical Model and Data Placement Strategies \\ for MEMS Storage Devices}

Yi-Reun Kim$^{*}$, Kyu-Young Whang$^{*}$, Min-Soo Kim$^{*}$, Il-Yeol Song$^{**}$ \\
\vspace*{-0.20cm}
$^{*}$Department of Computer Science \\
\vspace*{-0.20cm}
Korea Advanced Institute of Science and Technology\,(KAIST) \\
\vspace*{-0.20cm}
$^{**}$College of Information Science and Technology \\
\vspace*{-0.20cm}
Drexel University \\
\vspace*{-0.20cm}
e-mail:\,$^{*}$\{yrkim, kywhang, mskim\}@mozart.kaist.ac.kr, $^{**}$songiy@drexel.edu \\
\end{center}


\renewcommand{\baselinestretch}{1.6}
\begin{abstract}
{\small



MEMS storage devices are new non-volatile secondary storages that
have outstanding advantages over magnetic disks.
MEMS storage devices, however, are much different from magnetic
disks in the structure and access characteristics. They have
thousands of heads called {\it probe tips} and provide the following
two major access facilities: (1) {\it flexibility}\,: freely
selecting a set of probe tips for accessing data, (2) {\it
parallelism}\,: simultaneously reading and writing data with the set
of probe tips selected. Due to these characteristics, it is
nontrivial to find data placements that fully utilize the capability
of MEMS storage devices. In this paper, we propose a simple logical
model called the {\it Region-Sector}\,({\it RS}) model that
abstracts major characteristics affecting data retrieval
performance, such as flexibility and parallelism, from the physical
MEMS storage model. We also suggest heuristic data placement
strategies based on the RS model and derive new data placements for
relational data and two-dimensional spatial data by using those
strategies. Experimental results show that the proposed data
placements improve the data retrieval performance by up to
4.0\,times for relational data and by up to 4.8\,times for
two-dimensional spatial data of approximately 320\,Mbytes compared
with those of existing data placements. Further, these improvements
are expected to be more marked as the database size grows.

}
\end{abstract}
\renewcommand{\baselinestretch}{2.0}

\newtheorem{definition}{\bf Definition}
\newtheorem{strategy}{\bf Strategy}
\newtheorem{example}{\bf Example}
\newtheorem{property}{\bf Property}

\newenvironment{newidth}[2]{
 \begin{list}{}{
  \setlength{\topsep}{0pt}
  \setlength{\leftmargin}{#1}
  \setlength{\rightmargin}{#2}
  \setlength{\listparindent}{\parindent}
  \setlength{\itemindent}{\parindent}
  \setlength{\parsep}{\parskip}
 }
\item[]}{\end{list}}

%
%
\section{Introduction}
\vspace*{-0.50cm}

Micro-Electro-Mechanical Systems\,(MEMS) is a technology that
integrates electronic circuits and mechanical parts into one
chip\,\cite{Wise98}. MEMS storage devices are new non-volatile
secondary storages based on the MEMS technology. The prototypes of
MEMS storage devices have been developed by Carnegie Mellon
University\,(CMU), IBM laboratory, and Hewlett-Packard laboratory.
Recently, there have been a number of efforts to increase its
capacity and to improve the performance\,\cite{Hong06}.

MEMS storage devices have outstanding advantages compared with
magnetic disks: average access time is ten times faster, average
bandwidth is thirteen times larger, and power consumption is 54
times lower; their size is as small as $1cm^2$\,\cite{Sch00-ASPLOS}.
Due to these advantages, MEMS storage devices are expected to be
widely used in many places, such as the secondary storage of a
laptop\,\cite{Gri00-SIGMETRICS} and the middle-level storage to
reduce the performance gap between main memory and disk in the
memory hierarchy\,\cite{Ran03,Zhu04}.

MEMS storage devices, however, are much different from magnetic
disks in the structure and access characteristics. They have
thousands of heads called {\it probe tips} to access data. MEMS
storage devices also have the following two major access
characteristics\,\cite{Sch04-FAST}: (1) {\it flexibility}\,: freely
selecting a set of probe tips for accessing data, (2) {\it
parallelism}\,: simultaneously reading and writing data with the set
of probe tips selected. For good data retrieval performance, it is
necessary to place data on MEMS storage devices taking advantage of
their structures and access
characteristics\,\cite{Gri00-OSDI,Sch04-FAST,Yu06,Yu07,Zhu04}.

There have been a number of studies on data placement for MEMS
storage devices. In the operating systems field, methods have been
proposed that abstract the MEMS storage device as a linear array of
fixed-size logical blocks with one head\,\cite{Dra03,Gri00-OSDI}.
These methods allow us to use the MEMS storage device easily just
like a disk, but provide relatively poor data retrieval performance
because they do not take full advantage of the characteristics of
MEMS storage devices\,\cite{Sch04-FAST}. In the database field,
methods have been proposed to directly place data on the MEMS
storage device based on data access patterns of
applications\,\cite{Yu06,Yu07}. These methods provide relatively
good data retrieval performance\,\cite{Sch04-FAST}, but are quite
sophisticated because they directly manage MEMS storage devices
having a complicated structure.

In this paper, we propose a logical model called the {\it
Region-Sector}\,({\it RS}) model that abstracts the physical MEMS
storage model. The RS model abstracts major characteristics
affecting data retrieval performance -- flexibility and parallelism
-- from the physical MEMS storage model. The RS model is simple
enough for users to easily understand and use the MEMS storage
device and, at the same time, is strong enough to provide capability
comparable to that of a physical MEMS storage model. We also suggest
heuristic data placement strategies based on the RS model. These
strategies allow us to find data placements efficiently for a given
application.

The contributions of this paper are as follows: (1) we propose the
RS model, which is a logical abstraction of the MEMS storage device;
(2) we suggest heuristic data placement strategies based on the RS
model; (3) we derive new data placements for relational data and
two-dimensional spatial data by using those strategies; (4) through
extensive analysis and experiments, we show that the data retrieval
performances of our data placements are superior or comparable to
those of existing data placements.

The rest of this paper is organized as follows. Section\,2
introduces the MEMS storage device. Section\,3 describes prior art
related to data placement for the MEMS storage device. Section\,4
proposes the RS model. Section\,5 presents heuristic data placement
strategies. Section\,6 presents new data placements derived by using
heuristic data placement strategies. Section\,7 presents the results
of performance evaluation. Section\,8 summarizes and concludes the
paper.

%
%
\vspace*{0.65cm} 
\section{MEMS Storage Devices}
\vspace*{-0.30cm}

\begin{figure}[h!]
  \vspace*{0.50cm}
  \centerline{\psfig{file=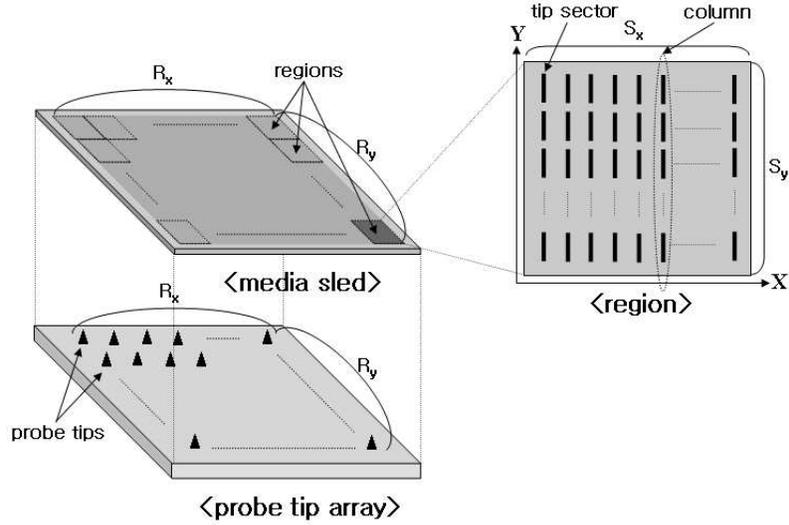, height=7cm}}
  \vspace*{-0.2cm}
  \caption{The structure of the MEMS storage device.}
  \label{fig:2-MEMS}
\end{figure}

The MEMS storage device is composed of a {\it media sled} and a {\it
probe tip array}. Figure~\ref{fig:2-MEMS} shows the structure of the
MEMS storage device. The {\it media sled} is a square plate on which
data is read and written by recording techniques such as magnetic,
thermomechanical, and phase-change ones\,\cite{Sch04-FAST}. The
media sled has $R_{x} \times R_{y}$ squares called {\it regions}.
Here, $R_{x}$\,($R_{y}$) is the number of regions in the X\,(Y)
axis. Each region contains $S_{x} \times S_{y}$ {\it tip sectors},
which are the smallest unit of accessing data. Here,
$S_{x}$\,($S_{y}$) is the number of tip sectors in a region in the
X\,(Y) axis. A {\it column} is a set of tip sectors that have the
same position in the X axis of each region\,\cite{Gri00-OSDI}. The
{\it probe tip array} is a set of $R_{x} \times R_{y}$ heads called
{\it probe tips}. Each probe tip reads and writes data on the
corresponding region of the media sled.

The MEMS storage device reads and writes data by moving the media
sled on the probe tip array. Here, a number of probe tips can be
activated so as to simultaneously read and write data. Each
activated probe tip reads or writes data on the tip sector having
the same relative position in each region. Users are able to freely
select a set of probe tips to be simultaneously activated, the
number of which is restricted to $200 \sim 2,000$ due to the
limitation in power consumption and electric
heat\,\cite{Gri00-SIGMETRICS}.

The major access characteristics\,\cite{Sch04-FAST} of the MEMS
storage device are summarized as follows.

\vspace*{-0.3cm}
\begin{description}
\item [Flexibility:] freely selecting and activating a set of probe tips for accessing data.

\item [Parallelism:] simultaneously reading and writing data with
the set of probe tips selected.
\end{description}

The MEMS storage device reads or writes data by performing the
following three steps\,\cite{Gri00-OSDI}.

\vspace*{-0.3cm}
\begin{description}
\item [1. Activating step:] activating a set of probe tips to use\,
(the activating time is negligible compared with seek or transfer
times).

\item [2. Seeking step:] moving the media sled so that the probe
tip is located on the target tip sector\,(the seek time is dependent
on the distance that the media sled moves).

\item [3. Transferring step:] reading or writing data on tip sectors that are contiguously
arranged within columns while moving the media sled in the +\,(or -)
direction of the Y axis\,(the transfer time is proportional to the
size of data accessed).

\vspace*{-0.3cm}
\end{description}

\noindent If tip sectors to be accessed are not contiguous within a
column but scattered over many columns, data are accessed by
performing the steps 2 and 3 repeatedly.

We explain the seek process in more detail since it is quite
different from that of the disk. The seek time can be computed using
Equations~(\ref{eq:eq-2-1})$\sim$(\ref{eq:eq-2-3}). Let
$SeekTime_{x}$ be the time to seek in the direction of the X axis,
and $SeekTime_{y}$ in the direction of the Y axis. In
$SeekTime_{x}$, if the media sled moves in the direction of the X
axis, we have to wait until the vibration of the media sled stops.
The time to wait for such vibration to stop is called the {\it
settle time}. Thus, $SeekTime_{x}$ is the sum of the move time and
the settle time as in Equation~(\ref{eq:eq-2-1}). In $SeekTime_{y}$,
if the media sled moves in the opposite direction of the current
direction, the media sled has to turn around. The time to turn
around is called the {\it turnaround time}. Thus, $SeekTime_{y}$ is
the sum of the move time and the turnaround time as in
Equation~(\ref{eq:eq-2-2}). If the media sled moves in the same
direction of the current direction, the turnaround time is zero.
Since the media sled is capable of moving in the direction of both
the X axis and the Y axis simultaneously, the total seek time is the
maximum of $SeekTime_{x}$ and $SeekTime_{y}$ as in
Equation~(\ref{eq:eq-2-3}).

\vspace*{-0.50cm}
\begin{eqnarray}
\label{eq:eq-2-1}
SeekTime_{x} & = & MoveTime_{x}~~~+~~~SettleTime \\
\label{eq:eq-2-2}
SeekTime_{y} & = & MoveTime_{y}~~~+~~~TurnaroundTime \\
\label{eq:eq-2-3} SeekTime     & = &
MAX~(~SeekTime_{x}\,,~~SeekTime_{y}~)
\end{eqnarray}

Table~\ref{tbl:mems} summarizes the parameters and values of the CMU
MEMS storage device being widely used for
research\,\cite{Car00b,Gri00-OSDI}. We use them in this paper. In
Table~\ref{tbl:mems}, $T_{X}$\,($T_{Y}$) is the average time to move
from one random position to another in the direction of the X\,(Y)
axis\,\cite{Car00b}.

\vspace*{-0.50cm} 
\vspace*{0.50cm}
\renewcommand{\baselinestretch}{1.10}
\begin{table}
\begin{center}
\caption{The parameters and values of the CMU MEMS storage device.}
\vspace*{0.3cm}
\begin{tabular} {|c|l|c|}
\hline
\multicolumn{1}{|c|}{Symbols} & \multicolumn{1}{c|}{Definitions} & \multicolumn{1}{c|}{Values} \\
\hline \hline
$R_{x}$ & the number of regions in the direction of the X axis & 80 \\
\hline
$R_{y}$ & the number of regions in the direction of the Y axis & 80 \\
\hline
$N_{R}$ & the number of regions (= $R_{x} \times R_{y}$) & $6,400$ \\
\hline
$S_{x}$ & the number of tip sectors in a region in the direction of the X axis & 2500 \\
\hline
$S_{y}$ & the number of tip sectors in a region in the direction of the Y axis & 27 \\
\hline
$N_{S}$ & the number of tip sectors in a region (= $S_{x} \times S_{y}$) & $67,500$ \\
\hline
$N_{PT}$ & the number of probe tips & 6,400 \\
\hline
$N_{APT}$ & the maximum number of active probe tips & 1,280 \\
\hline
$SectorSize$ & the size of data area in a tip sector (bits) & 64 \\
\hline
$TransferRate$ & the transfer rate per probe tip (Mbit/s) & 0.7 \\
\hline
$T_{X}$ & the average move time in the direction of the X axis (ms) & 0.52 \\
\hline
$T_{Y}$ & the average move time in the direction of the Y axis (ms) & 0.35 \\
\hline
$T_{S}$ & the average settle time (ms) & 0.215 \\
\hline
$T_{T}$ & the average turnaround time (ms) & 0.06 \\
\hline
\end{tabular}
\label{tbl:mems}
\end{center}
\end{table}
\renewcommand{\baselinestretch}{2.0}

%
%
\section{Related Work}
\vspace*{-0.30cm}

There have been a number of studies on data placement for the MEMS
storage device. We classify them into two categories -- {\it disk
mapping approaches} and {\it device-specific approaches} --
depending on whether they take advantage of the characteristics of
the storage device. This classification of the MEMS storage device
is analogous to that of the flash memory\,\cite{Gal05}, which is
another type of new non-volatile secondary storage. For the flash
memory, device-specific approaches\,(e.g., {\it Yet Another Flash
File System}\,({\it YAFFS})\,\cite{Man02}) provide new mechanisms to
exploit the features of the flash memory in order to improve
performance, while disk mapping approaches(e.g., {\it Flash
Translation Layer}\,({\it FTL})\,\cite{Ban95}) abstract the flash
memory as a linear array of fixed-size pages in order to use
existing disk-based algorithms on the flash memory. In this section,
we explain two categories for the MEMS storage device in more
detail.

\vspace*{0.10cm} 
\vspace*{-0.1cm}
\subsection{Disk Mapping Approaches}
\vspace*{-0.1cm}

Griffin et al.\,\cite{Gri00-OSDI} and Dramaliev et al.\,\cite{Dra03}
proposed models to use the MEMS storage device just like a disk.
They abstract the MEMS storage device as a linear array of
fixed-size logical blocks with one head. This linear abstraction
works well for most applications using the MEMS storage device as
the replacement of the disk\,\cite{Gri00-OSDI}. However, they
provide relatively poor data retrieval performance compared with
device-specific approaches\,\cite{Yu06,Yu07} because they do not
take full advantage of the characteristics of the MEMS storage
device\,\cite{Sch04-FAST}.

\vspace*{0.10cm} 
\vspace*{-0.1cm}
\subsection{Device-specific Approaches}
\vspace*{-0.1cm}

Yu et al.\,\cite{Yu06,Yu07} proposed methods for placing data on the
MEMS storage device based on data access patterns of applications.
Yu et al.\,\cite{Yu07} places relational data on the MEMS storage
device such that projection queries are performed efficiently. Yu et
al.\,\cite{Yu06} places two-dimensional spatial data such that
spatial range queries are performed efficiently. These data
placements identify that data access patterns of such applications
are inherently two-dimensional, and then, place data so as to take
advantage of parallelism and flexibility of the MEMS storage device.
We explain each data placement in more detail for comparing them
with our methods in Section\,6.

\vspace*{-0.1cm}
\subsubsection{Data Placement for Relational Data}
\vspace*{-0.1cm}

\begin{figure}[h!]
  \vspace*{0.50cm}
  \centerline{\psfig{file=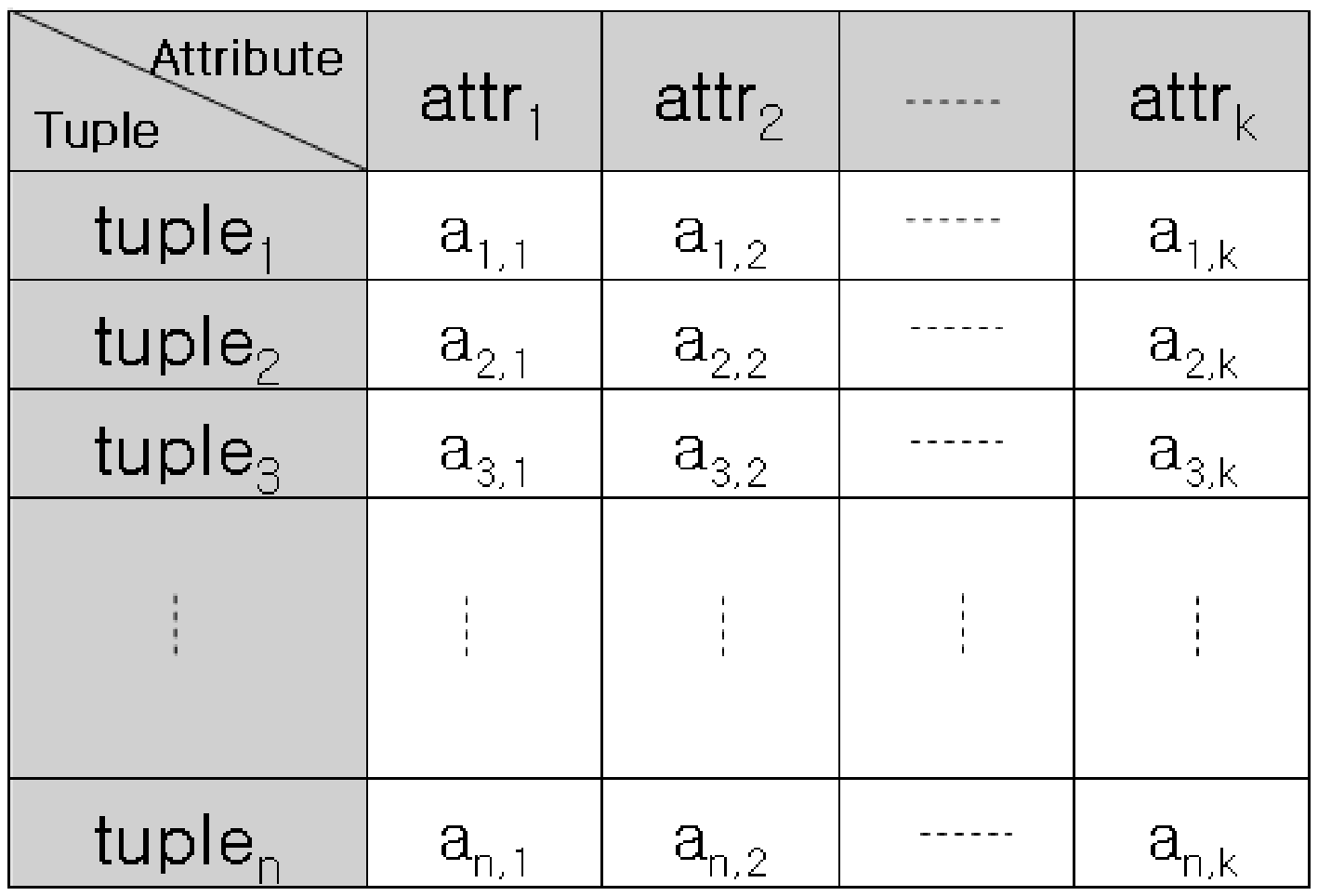, height=4cm}}
  \vspace*{-0.2cm}
  \caption{An example relation $R$.}
  \label{fig:5-RelationalTable}
\end{figure}

Yu et al.\,\cite{Yu07} deals with the application that places a
relation on the MEMS storage device, and then, executes simple
projection queries over that relation. Here, queries read the values
of the specified attributes of all tuples.
Figure~\ref{fig:5-RelationalTable} shows an example relation $R$,
which has $k$ attributes $attr_{1}, ..., attr_{k}$ and has $n$
tuples. Here, $a_{i,j}$ represents the $j$\,th attribute value of
the $i$\,th tuple\,($1 \leq i \leq n$, $1 \leq j \leq k$).

\begin{figure}[h!]
  \vspace*{-0.70cm} 
  \vspace*{0.50cm}
  \centerline{\psfig{file=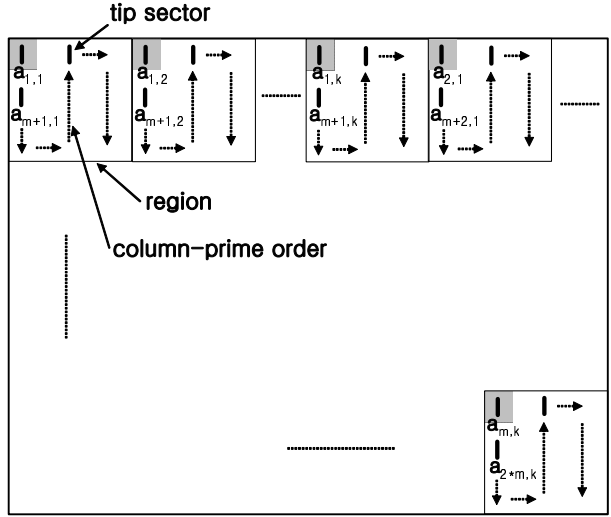, height=7cm}}
  \vspace*{-0.2cm}
  \caption{Yu et al.'s data placement.}
  \label{fig:5-RelationalTable-Yu-MEMS}
\end{figure}

Figure~\ref{fig:5-RelationalTable-Yu-MEMS} shows Yu et
al.\,\cite{Yu07}'s data placement of the relation $R$ on the MEMS
storage device. Here, for simplicity of explanation, we assume that
the length of each attribute value is equal to the size of the tip
sector. First, a set of $m$ tuples ($tuple_{1} \sim tuple_{m}$) is
placed on the first tip sector of each region, i.e., the shaded tip
sectors in Figure~\ref{fig:5-RelationalTable-Yu-MEMS}. Likewise,
each set of $m$ tuples ($tuple_{m \times (i-1) + 1} \sim tuple_{m
\times i}$) is placed on the $i$\,th tip sector of the region\,($2
\leq i \leq \lceil \frac{n}{m} \rceil$) in the column-prime order.
Equation~(\ref{eq:eq-5-1-2}) shows a mapping function
$f_{RelationtoMEMS}$ that puts the attribute value $a_{v,w}$ into
the tip sector $<$$r_{x}, r_{y},s_{x},s_{y}$$>$ of the MEMS storage
device.

\clearpage 

\begin{figure}[h!]
  \vspace*{-0.5cm}
  \begin{eqnarray}
  f_{RelationtoMEMS}\,(a_{v,w}) = \left\{ \begin{array}{l}
          r_{x} = ((k \times ((v-1)~\mathtt{mod}~m)+w)-1)~\mathtt{mod}~R_{x} + 1 \\
          r_{y} = \lceil \frac{(k \times ((v-1)\,\mathtt{mod}\,m)+w)}{R_{x}} \rceil \\
          s_{x} = \lceil \frac{\lceil \frac{v}{m} \rceil}{S_{y}} \rceil \\
          s_{y} = \left\{ \begin{array}{ll} (\lceil \frac{v}{m} \rceil-1)~\mathtt{mod}~S_{y} + 1           & \mbox{if $s_{x}$ is odd} \\
                                            S_{y} - ((\lceil \frac{v}{m} \rceil-1)~\mathtt{mod}~S_{y}) & \mbox{if $s_{x}$ is even}
  \end{array} \right.
  \end{array} \right.
  \label{eq:eq-5-1-2}
  \end{eqnarray}
\end{figure}

\vspace*{-0.40cm} 
\vspace*{-0.1cm}
\subsubsection{Data Placement for Two-dimensional Spatial Data}
\vspace*{-0.1cm}

\begin{figure}[h!]
  \vspace*{0.50cm}
  \centerline{\psfig{file=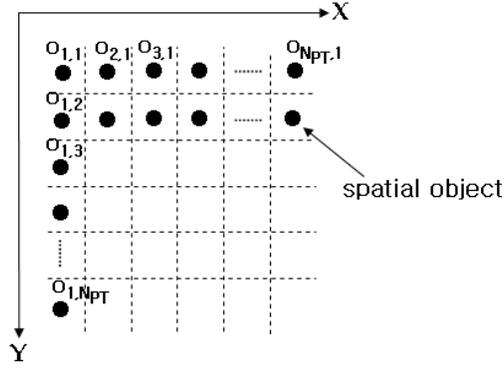, height=5cm}}
  \vspace*{-0.2cm}
  \caption{An example set $S$ of two-dimensional spatial objects.}
  \label{fig:5-SpatialData}
\end{figure}

Yu et al.\,\cite{Yu06} deals with an application that places a set
of two-dimensional spatial objects on the MEMS storage device, and
then, executes region queries over those objects. Here, the
two-dimensional spatial objects are uniformly distributed in the
two-dimensional space, and queries read objects contained in a
rectangular region. Figure~\ref{fig:5-SpatialData} shows an example
set $S$ of two-dimensional $N_{PT} \times N_{PT}$ spatial objects.

\begin{figure}[h!]
  \vspace*{0.50cm}
  \centerline{\psfig{file=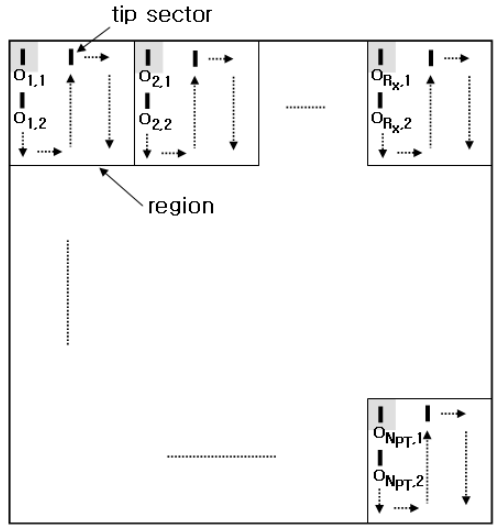, height=7cm}}
  \vspace*{-0.2cm}
  \caption{Yu et al.'s data placement.}
  \label{fig:5-SpatialData-Yu-MEMS}
\end{figure}

Figure~\ref{fig:5-SpatialData-Yu-MEMS} shows Yu et
al.\,\cite{Yu06}'s data placement of the set $S$ in the MEMS storage
device. Here, for simplicity of explanation, we assume that each
object is stored in one tip sector. In
Figure~\ref{fig:5-SpatialData-Yu-MEMS}, the objects from $o_{1,1}$
to $o_{N_{PT},1}$ are first placed on the first tip sector of each
region. Likewise, the objects from $o_{1,i}$ to $o_{N_{PT},i}$ on
the $i$\,th tip sector of each region\,($2 \leq i \leq 6400$) in the
column-prime order. Equation~(\ref{eq:eq-5-2-2}) shows a mapping
function $f_{SpacetoMEMS}$ that places the object $o_{x,y}$ on the
tip sector $<$$r_{x}, r_{y},s_{x},s_{y}$$>$ of the MEMS storage
device.

\begin{figure}[h!]
  \vspace*{-0.5cm}
  \begin{eqnarray}
  f_{SpacetoMEMS}\,(o_{x,y}) = \left\{ \begin{array}{l}
          r_{x} = (x-1)~\mathtt{mod}~N_{R_{x}} + 1 \\
          r_{y} = \lceil \frac{x}{N_{R_{x}}} \rceil \\
          s_{x} = \lceil \frac{y}{N_{S_{y}}} \rceil \\
          s_{y} = \left\{ \begin{array}{ll} (y-1)~\mathtt{mod}~N_{S_{y}} + 1           & \mbox{if $s_{x}$ is odd} \\
                                            N_{S_{y}} - ((y-1)~\mathtt{mod}~N_{S_{y}}) & \mbox{if $s_{x}$ is even}
  \end{array} \right.
  \end{array} \right.
  \label{eq:eq-5-2-2}
  \end{eqnarray}
\end{figure}

\clearpage 

%
%
\section{Region-Sector\,(RS) Model for the MEMS Storage Device}
\vspace*{-0.30cm}

In this Section, we propose the RS model for the MEMS storage
device. In Section\,4.1, we provide an overview of the RS model. In
Section\,4.2, we formally define the RS model. In Section\,4.3, we
present the mapping function between the RS model and the MEMS
storage device.

\vspace*{-0.1cm}
\subsection{Overview}
\vspace*{-0.1cm}

The RS model can be regarded as a {\it virtual view} of the physical
MEMS storage device. The purpose of the model is to provide an
abstraction making it easy to understand and simple to use the
complex MEMS storage device while maintaining its performance and
flexibility.

When placing data on the disk, the OS and applications abstract the
disk as a relatively simple logical view such as a linear array of
fixed-sized logical blocks because considering the physical
structures\,(cylinders, tracks, and sectors) of the disk is complex.
This kind of abstraction can also be applied to the MEMS storage
device. By abstracting the MEMS storage device as a relatively
simple logical view such as the RS model, we can more easily place
data on the MEMS storage device than when we directly consider the
physical structures\,(regions, columns, tip sectors).

\begin{figure}[h]
  \vspace*{0.50cm}
  \centerline{\psfig{file=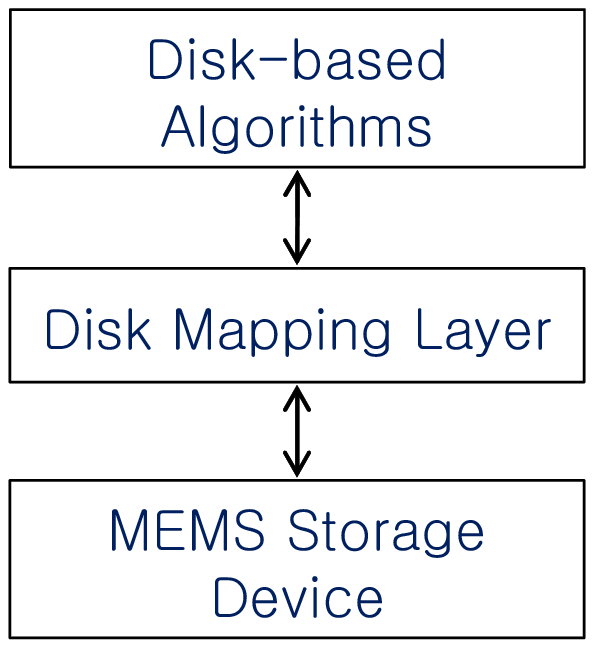, width=4cm}
              \hspace*{1.50cm}
              \psfig{file=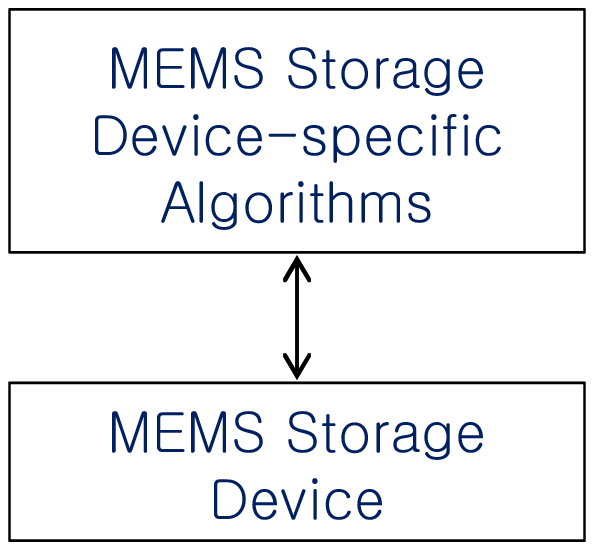, width=4cm}
              \hspace*{1.50cm}
              \psfig{file=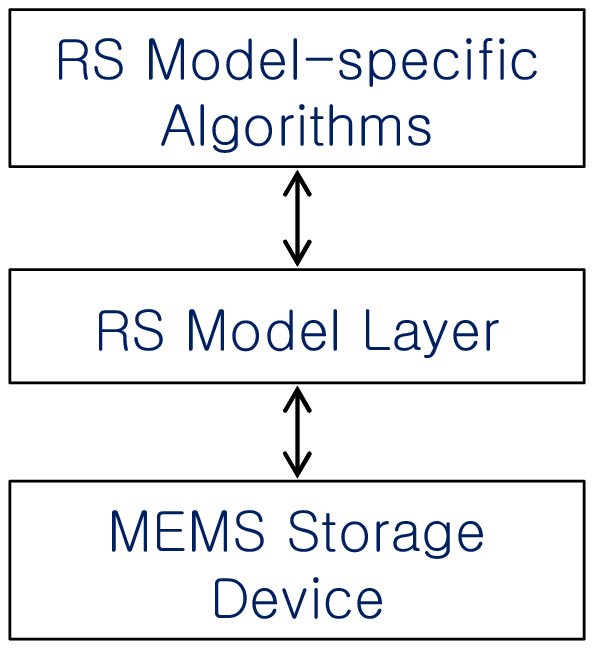, width=4cm}}
  \centerline{(a) The disk mapping layer \hspace*{1.0cm} (b) The device-specific algorithm \hspace*{0.8cm} (c) The RS model layer \hspace*{0.3cm} }
  \vspace*{-0.2cm}\centerline{architecture. \hspace*{3.5cm} architecture. \hspace*{3.5cm} architecture.}
  \vspace*{-0.2cm}
  \caption{The architectures of the system for the MEMS storage device.}
  \label{fig:3-SystemArchitecture}
\end{figure}

Figure~\ref{fig:3-SystemArchitecture} shows three kinds of system
architectures for using the MEMS storage device.
Figure~\ref{fig:3-SystemArchitecture}(a) shows one using the
disk-based algorithms and the disk mapping layer\,(explained in
Section 3.1); Figure~\ref{fig:3-SystemArchitecture}(b) one using the
MEMS storage device-specific algorithms\,(explained in Section 3.2)
without any mapping layer; and
Figure~\ref{fig:3-SystemArchitecture}(c) one using the RS
model-specific algorithms and the RS model layer. The architecture
in Figure~\ref{fig:3-SystemArchitecture}(c) is capable of providing
higher performance compared with that in
Figure~\ref{fig:3-SystemArchitecture}(a) by taking advantage of
useful characteristics of the MEMS storage device through the RS
model. It also helps us find good data placements for a given
application more easily than the architecture in
Figure~\ref{fig:3-SystemArchitecture}(b) because it hides complex
features of the physical MEMS storage device.

\vspace*{-0.1cm}
\subsection{Definition of the RS Model}
\vspace*{-0.1cm}

The RS model maps the tip sectors of the MEMS storage device into a
virtual two-dimensional plane in order to effectively use
parallelism and flexibility. For the mapping, we first classify the
tip sectors into two groups depending on the possibility of using
parallelism. It is possible to use parallelism for the tip sectors
having the same relative $(x,y)$ position in each region because we
are able to freely select a set of tip sectors and simultaneously
access them. Hereafter, we call the set of tip sectors having the
same relative $(x,y)$ positions in each region as the {\it
simultaneous-access sector group}. On the other hand, it is not
possible to use parallelism for the tip sectors existing in the same
region because we are able to access only one tip sector at a time
from them. Hereafter, we call the set of such tip sectors as the
{\it non-simultaneous-access sector group}.

\begin{figure}[h!]
  \vspace*{0.50cm}
  \centerline{\psfig{file=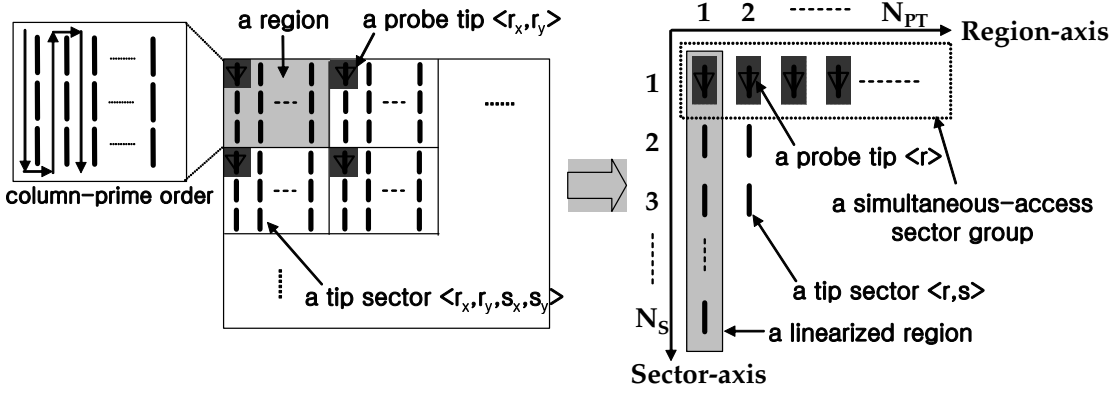, height=5.5cm}}
  \centerline{\hspace*{1.0cm} (a) The MEMS storage device. \hspace*{3.0cm} (b) The Region-Sector\,(RS) model.}
  \vspace*{-0.2cm}
  \caption{The structure of the RS model.}
  \label{fig:3-RegionSectorModel}
\end{figure}

Figure~\ref{fig:3-RegionSectorModel} shows the structure of the RS
model. The RS model is composed of a set of probe tips and a
two-dimensional plane. The set of probe tips are lined up
horizontally. We call them the {\it probe tip line}. The
two-dimensional plane has the {\it Region} axis and the {\it Sector}
axis. The RS model maps the tip sectors in a simultaneous-access
sector group in the direction of the Region axis and those in a
non-simultaneous-access sector group in the direction of the Sector
axis. We map the tip sectors in the non-simultaneous-access sector
group\,(i.e., tip sectors in a region) in the {\it
column-prime\,order} as shown in
Figure~\ref{fig:3-RegionSectorModel} since it is the fastest order
to access all the tip sectors in a region\,\cite{Sch00-ASPLOS,Yu07}.
We call an ordered set of tip sectors that have the same position in
the Region axis a {\it linearized region}. The RS model regards the
tip sectors within a linearized region as {\it quasi-contiguous}.
Each probe tip reads and writes data on the corresponding linearized
region of the RS model.

The RS model simplifies the structure of the MEMS storage device by
reducing the number of parameters to represent the position of a tip
sector. In the MEMS storage device, the position of a tip sector is
represented by four parameters $<$$r_x, r_y, s_x,
s_y$$>$\,($1$$\leq$$r_x$$\leq$$R_x$, $1$$\leq$$r_y$$\leq$$R_y$,
$1$$\leq$$s_x$$\leq$$S_x$, $1$$\leq$$s_y$$\leq$$S_y$) as shown in
Figure~\ref{fig:3-RegionSectorModel}(a), where $<$$r_x, r_y$$>$ is
the position of the region and $<$$s_x, s_y$$>$ the position of the
tip sector within the region. On the other hand, in the RS model,
the position of a tip sector is represented by only two parameters
$<$$r, s$$>$ as shown in Figure~\ref{fig:3-RegionSectorModel}(b),
where $r$ is the position of the tip sector in the Region axis and
$s$ in the Sector axis.

The RS model reads or writes data by performing the following three
steps repeatedly\,(as compared to the physical MEMS storage device
described in Section\,2).

\vspace*{-0.3cm}
\begin{description}
\item [1. Activating step:] activating a set of probe tips to use.

\item [2. Seeking step:] moving the probe tip line to the target row.

\item [3. Transferring step:] reading or writing data on tip sectors
that are quasi-contiguously arranged within linearized regions while
moving the probe tip line in the +\,(or -) direction of the Sector
axis.

\vspace*{-0.3cm}
\end{description}

\noindent The RS model considers quasi-contiguous tip sectors within
a linearized region to be sequentially accessed\,(the reason will be
explained later), while the MEMS storage device is capable of
sequentially accessing contiguous tip sectors only within a column.

We explain the seek time and transfer rate of the RS model. Through
calculation using them, users can approximately estimate the data
access time in the MEMS storage device exactly mapping the data to
the MEMS storage device. The calculation of data access time in the
RS model is easier because the movement of probe tips in the RS
model is modeled simpler than that in the MEMS storage device.

For the seek time of the RS model, for simplicity, we use the
average seek time of the physical MEMS storage device. By using the
average seek time instead of the real seek time, we can
significantly simplify the cost model for data retrieval performance
while little sacrificing the accuracy of the cost model.

In the RS model, the transfer rate per probe tip is calculated as
the data size of a region divided by the time to read all the tip
sectors of a region in the column-prime order. We note that the RS
model considers all quasi-contiguous tip sectors within a linearized
region to be sequentially accessed. Table~\ref{tbl:transferrate-rs}
summarizes some notation to be used for calculating the transfer
rate.

\vspace*{0.50cm}
\renewcommand{\baselinestretch}{1.10}
\begin{table}
\begin{center}
\caption{The notation to be used for calculating the transfer rate
per probe tip in the RS model.} \vspace*{0.3cm}
\begin{tabular} {|c|l|}
\hline
\multicolumn{1}{|c|}{Symbols} & \multicolumn{1}{c|}{Definitions} \\
\hline \hline
$S_{x}$ & the number of columns in a region \\
\hline
$S_{y}$ & the number of tip sectors in a column \\
\hline
$SectorSize$ & the size of a tip sector (bytes) \\
\hline
$RegionSize$ & the size of a region (bytes) (= $S_{x} \times S_{y} \times SectorSize$) \\
\hline
$TransferRate$ & the transfer rate per probe tip in the physical MEMS storage device (Mbytes/s) \\
\hline
$SeekTime_{adj}$ & the seek time from a column to an adjacent column in the physical MEMS \\
                 & storage device (s) \\
\hline
\end{tabular}
\label{tbl:transferrate-rs}
\end{center}
\end{table}
\renewcommand{\baselinestretch}{2.0}

The transfer rate per probe tip in the RS model is computed as in
Equation~(\ref{eq:eq-3-2}). The time to read data of a region in the
column-prime order is the sum of the following two terms: (1) the
time to read data of each column, (2) the time to seek to the
adjacent column for each column.
The former is $\frac{RegionSize}{TransferRate}$, and the latter
$S_{x} \times SeekTime_{adj}$. $SeekTime_{adj}$ is computed as in
Equation~(\ref{eq:eq-3-3}). Because the move time to the adjacent
column $MoveTime_{adj\_x}$ is negligible compared with $SettleTime$,
and $SettleTime$ is larger than $TurnaroundTime$, $SeekTime_{adj}$
is approximately equal to $SettleTime$.

\vspace*{-0.5cm}
\begin{eqnarray}
TransferRate_{rs} & = & \frac{RegionSize}
                             {(\,\frac{RegionSize}{TransferRate}\,)~~~+~~~(\,S_{x} \times {SeekTime_{adj}}\,)}
\label{eq:eq-3-2}
\end{eqnarray}

\vspace*{-0.5cm}
\begin{eqnarray}
SeekTime_{adj} & =    & MAX\,(\,MoveTime_{adj\_x}~+~SettleTime\,,~~TurnaroundTime\,) \nonumber \\
               & \approx & SettleTime
\label{eq:eq-3-3}
\end{eqnarray}

The characteristics of the RS model in both random and sequential
accesses are not much different from those of the MEMS storage
device. The seek time of the RS model is equal to that of the MEMS
storage device since the RS model uses the average time to seek from
one random position to another in a certain region of the MEMS
storage device. In Equation~(\ref{eq:eq-3-2}), the total seek
time\,(i.e., $S_{x} \times SeekTime_{adj}$) is only about 6\,\% of
the time to read all the tip sectors of a region. Thus, the transfer
rate of the RS model is approximately equal to that of the MEMS
storage device.

Table~\ref{tbl:comparision-RSmodel} summarizes the differences
between the RS model and the physical MEMS storage model.

\vspace*{0.50cm}
\renewcommand{\baselinestretch}{1.10}
\begin{table}
\begin{center}
\caption{Comparison of the RS model with the physical MEMS storage
model.} \vspace*{0.3cm}
\begin{tabular} {|c|c|c|c|}
\hline
\multicolumn{1}{|c|}{} & \multicolumn{1}{c|}{MEMS storage model} & \multicolumn{1}{c|}{RS model} & \multicolumn{1}{c|}{Remarks}\\
\hline \hline
addressing the position & $<$$r_x, r_y, s_x, s_y$$>$ & $<$$r, s$$>$                       & simpler    \\
of a tip sector         &                            &                                    &             \\
\hline
movement of             & in the +/- direction of    & in the +/- direction of            & simpler     \\
probe tips              & the X and Y axes           & the Sector axis                    &             \\
\hline
the area of             & $S_y$ tip sectors          & $N_S = S_x \times S_y$ tip sectors & expanded by $S_x$ times \\
sequential access       & within a column            & within a linearized region         & \\
                        &                            & (quasi-contiguous)                 & \\
\hline
seek time               & real seek time             & average seek time                  & equal in average \\
                        &                            & from one random position           & \\
                        &                            & to another                         & \\
\hline
transfer rate           & real transfer rate         & average transfer rate              & approximately equal \\
                        &                            & when accessing tip sectors         & \\
                        &                            & in a region in the                 & \\
                        &                            & column-prime order                 & \\
\hline
\end{tabular}
\label{tbl:comparision-RSmodel}
\end{center}
\end{table}
\renewcommand{\baselinestretch}{2.0}

\vspace*{-0.1cm}
\subsection{Mapping Functions between the RS Model and the MEMS Storage Device}
\vspace*{-0.1cm}

In order to use the RS model, it is necessary to map the position of
each tip sector in the RS model into that in the MEMS storage model,
and {\it vice versa}. In this section, we define two mapping
functions $f_{RStoMEMS}$ and $f_{MEMStoRS}$. In
Equation~(\ref{eq:eq-3-4}), $f_{RStoMEMS}$ is for converting the
position $<$$r, s$$>$ in the RS model into the position $<$$r_{x},
r_{y},s_{x},s_{y}$$>$ in the MEMS storage model. In
Equation~(\ref{eq:eq-3-5}), $f_{MEMStoRS}$ is for converting the
position $<$$r_{x},r_{y},s_{x},s_{y}$$>$ into the position $<$$r,
s$$>$.

\begin{eqnarray}
f_{RStoMEMS}\,(<r,~s>) = \left\{ \begin{array}{l}
        r_{x} = (r-1)~\mathtt{mod}~R_{x} + 1 \\
        r_{y} = \lfloor \frac{(r-1)}{R_{x}} \rfloor + 1 \\
        s_{x} = \lfloor \frac{(s-1)}{S_{y}} \rfloor + 1 \\
        s_{y} = \left\{ \begin{array}{ll} (s-1)~\mathtt{mod}~S_{y} + 1& \mbox{if $s_{x}$ is odd} \\
                                          S_{y} - ((s-1)~\mathtt{mod}~S_{y}) & \mbox{if $s_{x}$ is even}
\end{array} \right.
\end{array} \right.
\label{eq:eq-3-4}
\end{eqnarray}

\vspace*{-0.5cm}
\begin{eqnarray}
f_{MEMStoRS}\,(<r_{x},~r_{y},~s_{x},~s_{y}>) = \left\{
\begin{array}{l}
        r = (R_{x} \times (r_{y}-1)) + r_{x} \\
        s = \left\{ \begin{array}{ll} (S_{y} \times (s_{x}-1)) + s_{y}& \mbox{if $s_{x}$ is odd} \\
                                      (S_{y} \times (s_{x}-1)) + (S_{y} - s_{y} + 1) & \mbox{if $s_{x}$ is even}
\end{array} \right.
\end{array} \right.
\label{eq:eq-3-5}
\end{eqnarray}

In practice, two mapping functions $f_{RStoMEMS}$ and $f_{MEMStoRS}$
are implemented as a driver between user algorithms\,(i.e., RS
model-specific algorithms in
Figure~\ref{fig:3-SystemArchitecture}(c)) and the MEMS storage
device. If users write and execute programs that place and access
data on the RS model, the data are automatically placed and accessed
on the MEMS storage device by this driver.

%
%
\section{Data Placement Strategies in the RS model}
\vspace*{-0.30cm}

For secondary storage devices, data retrieval performance is
significantly affected by data placement on them. The same holds for
the MEMS storage device. For good data retrieval performance, we
need to place data on the MEMS storage device taking advantage of
its structure and access
characteristics\,\cite{Gri00-OSDI,Sch04-FAST,Yu06,Yu07,Zhu04}. In
this section, we present heuristic data placement strategies that
help us efficiently find good data placements.

As the measure of data retrieval performance, we use the time to
read the data being retrieved by a query as was done by Yu et
al.\,\cite{Yu06,Yu07}. We call it the {\it retrieval time}.
Table~\ref{tbl:queryexeciotime} summarizes the notation to be used
for analyzing the retrieval time in the RS model.

\vspace*{0.50cm}
\renewcommand{\baselinestretch}{1.10}
\begin{table}
\begin{center}
\caption{The notation to be used for analyzing the retrieval time in
the RS model.} \vspace*{0.3cm}
\begin{tabular} {|c|l|}
\hline
\multicolumn{1}{|c|}{Symbols} & \multicolumn{1}{c|}{Definitions} \\
\hline \hline
$ RetrievalDataSize $ & the size of the data being retrieved by a query (bytes) \\
\hline
$ TransferRate_{rs} $ & the average transfer rate per probe tip in the RS model (Mbytes/s) \\
\hline
$ SeekTime_{rs} $ & the average seek time in the RS model (s) \\
\hline
$ K_{parallel} $ & the average number of probe tips used during query processing \\
\hline
$ K_{random} $ & the average number of seek operations occurring during query processing \\
\hline
\end{tabular}
\label{tbl:queryexeciotime}
\end{center}
\end{table}
\renewcommand{\baselinestretch}{2.0}

The retrieval time in the RS model can be computed as in
Equation~(\ref{eq:eq-4-1}). It is the sum of $Total\-TransferTime$
and $TotalSeekTime$. $TotalTransfer\-Time$ is $RetrievalDataSize$
divided by the total transfer rate, which is $TransferRate_{rs}
\times K_{parallel}$. $TotalSeekTime$ is $SeekTime_{rs} \times
K_{random}$.

\vspace*{-0.50cm}
\begin{eqnarray}
RetrievalTime & = & TotalTransferTime~~~+~~~TotalSeekTime \nonumber \\
                & = & (\,\frac{RetrievalDataSize}{TransferRate_{rs} \times K_{parallel}}\,)~~~+~~~
                      (\,SeekTime_{rs} \times K_{random}\,)
\label{eq:eq-4-1}
\end{eqnarray}

From Equation~(\ref{eq:eq-4-1}), we know that $RetrievalTime$
decreases as $K_{parallel}$ gets larger and as $K_{random}$ gets
smaller. Thus, for good performance, it is preferable to place data
such that $K_{parallel}$ is made as large as possible\,(its maximum
value is $N_{APT}$) and $K_{random}$ as small as possible\,(its
minimum value is $0$). Theoretically, the data placement that makes
$K_{parallel} = N_{APT}$ and, at the same time, $K_{random} = 0$ is
the optimal. However, it may not be feasible to find such data
placements. Hence, we employ two simple heuristic data placement
strategies as follows.

\vspace*{-0.3cm}
\begin{description}
\item [Strategy\_Sequential:] a strategy that
places the data being retrieved by a query as contiguously as
possible in the direction of the Sector axis in the RS model. This
strategy aims at making $K_{random}$ be as close to $0$ as possible.

\item [Strategy\_Parallel:] a strategy that places the
data being retrieved by a query as widely as possible in the
direction of the Region axis on the RS model. This strategy aims at
making $K_{parallel}$ be as close to $N_{APT}$ as possible.

\vspace*{-0.3cm}
\end{description}

%
%
\section{Applications of Data Placement Strategies}
\vspace*{-0.30cm}

In this Section, we present data placements derived from
Strategy\_Sequential and Strategy\_Parallel for two applications. We
present data placements for relational data in Section\,6.1, and
data placements for two-dimensional spatial data in Section\,6.2.

\vspace*{-0.1cm}
\subsection{Data Placements for Relational Data}
\vspace*{-0.1cm}

In this section, we deal with an application that places a relation
on the MEMS storage device, and then, executes simple projection
queries over that relation. This application is the same one dealt
with by Yu et al.\,\cite{Yu07} as described in Section\,3.2.1. We
present two data placements for relational data. We name the data
placement derived from Strategy\_Sequential, which turns out to be
identical to the placement proposed by Yu et al.\,\cite{Yu07}, as
{\it Relational-Sequential-Yu}, and the one derived from
Strategy\_Parallel as {\it Relational-Parallel}.

\vspace*{-0.1cm}
\subsubsection{Relational-Sequential-Yu}
\vspace*{-0.1cm}

Relational-Sequential-Yu intends to provide highly sequential
reading of data by preventing seek operations in processing queries.
Here, it is preferable that the values of the projected attributes
are placed as contiguously as possible in the direction of the
Sector axis. Accordingly, Relational-Sequential-Yu stores the tuples
of the relation $R$ such that a linearized region is occupied with
the values of only one attribute. Thus, these values are stored
quasi-contiguously.

\begin{figure}[h!]
  \vspace*{0.50cm}
  \centerline{\psfig{file=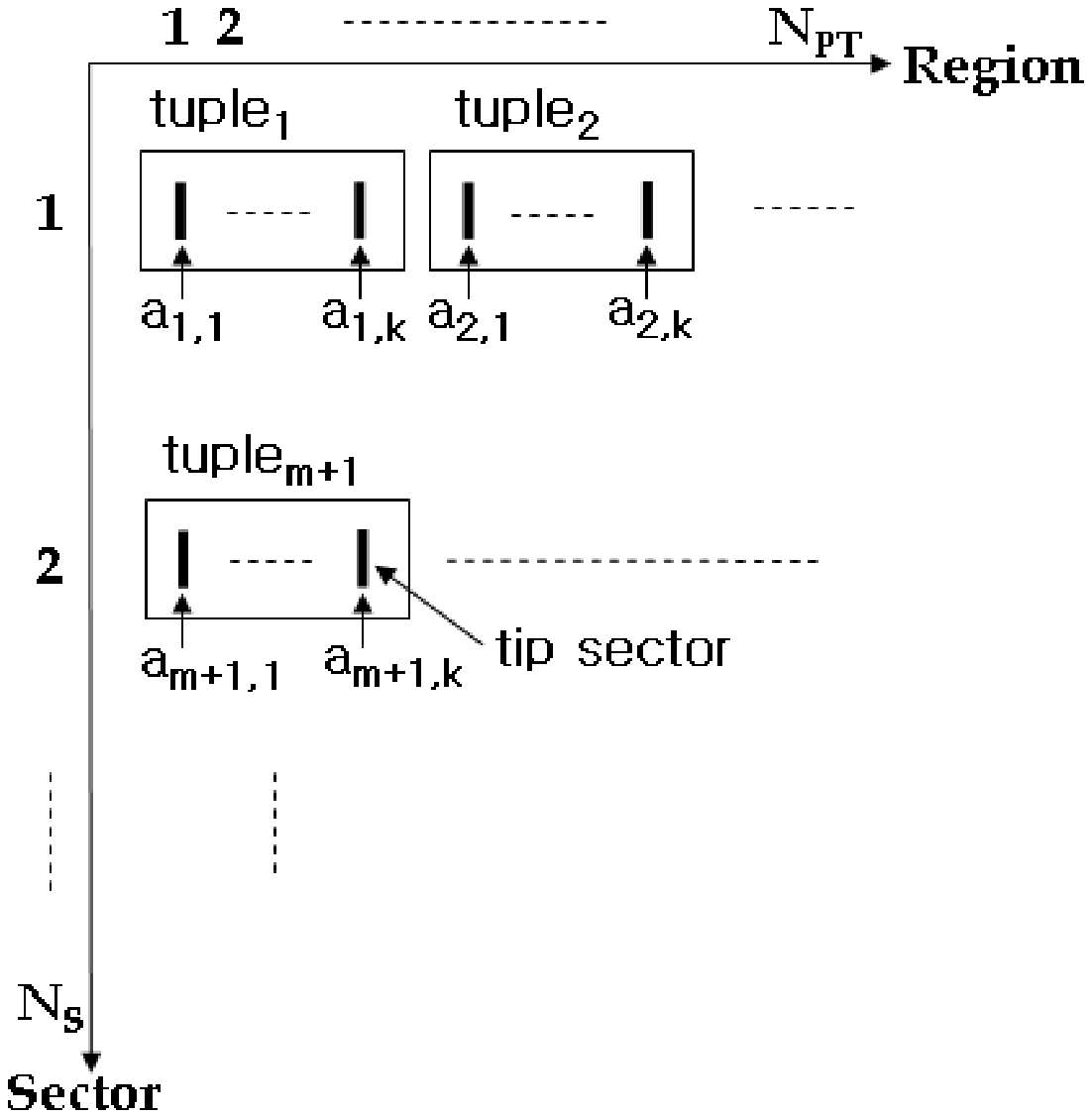, height=5.5cm}
              \hspace*{2.0cm}
              \psfig{file=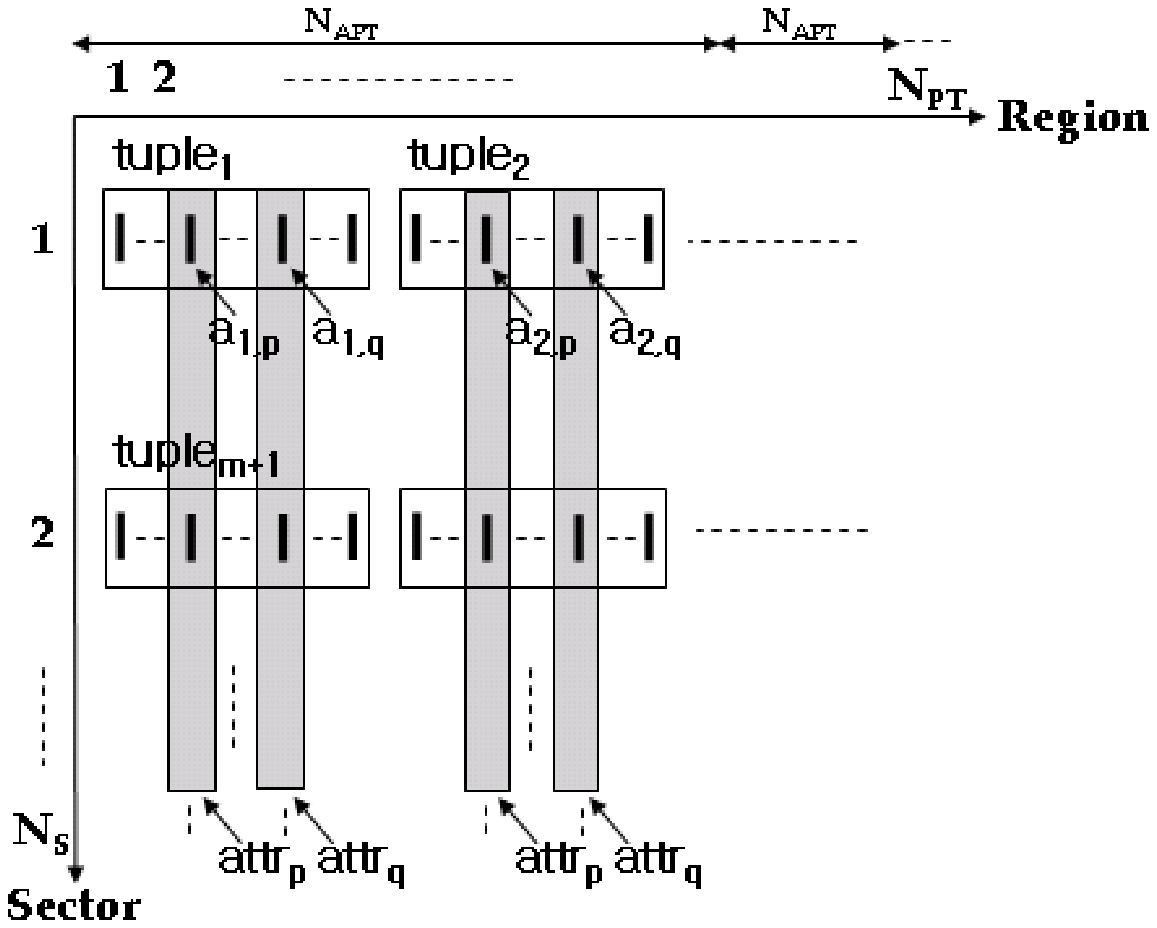, height=5.5cm}}
  \centerline{(a) Relational-Sequential-Yu. \hspace*{2.0cm} (b) The data area being retrieved by a query. \hspace*{0.0cm}}
  \vspace*{-0.2cm}
  \caption{Relational-Sequential-Yu data placement and the data area being retrieved by the query projecting $attr_{p}$ and $attr_{q}$.}
  \label{fig:5-RelationalTable-Yu}
\end{figure}

Figure~\ref{fig:5-RelationalTable-Yu} shows Relational-Sequential-Yu
and the data area being retrieved by the query projecting
$N_{projection}$ attributes. Let us assume that at most $m$ tuples
are stored in one simultaneous-access sector group. As shown in
Figure~\ref{fig:5-RelationalTable-Yu}(a), Relational-Sequential-Yu
puts $m$ tuples $tuple_{m \times (i-1) + j}$\,($1 \leq j \leq m$)
into the $i$\,th simultaneous-access sector group\,($1 \leq i \leq
\lceil \frac{n}{m} \rceil$). Equation~(\ref{eq:eq-5-1-1}) shows the
mapping function $f_{RelationtoRS}$ that puts the attribute value
$a_{v,w}$ into the tip sector $<$$r, s$$>$ in the RS model. In
Figure~\ref{fig:5-RelationalTable-Yu}(b), the shaded area indicates
the tip sectors accessed by the query projecting $attr_{p}$ and
$attr_{q}$. If the width of the shaded area\,(i.e., the number of
tip sectors corresponding to $attr_{p}$ or $attr_{q}$ in a
simultaneous-access sector group $= m \times N_{projection}$) is
less than or equal to $N_{APT}$, only one sequential scan suffices
for query processing. Otherwise, several sequential scans\,($=
\lceil \frac{m \times N_{projection}}{N_{APT}} \rceil$) are
required. We use column-prime order among scans by activating
another set of $N_{APT}$ probe tips\,\footnote{\vspace*{-0.2cm} For
each scan, a turnaround operation occurs in practice. But, the
turnaround operation is not a seek operation, and the time is
negligible compared with seek time or transfer time.}.

\begin{figure}[h!]
  \vspace*{-0.5cm}
  \begin{eqnarray}
  f_{RelationtoRS}\,(a_{v,w}) = \left\{ \begin{array}{l}
          r = k \times ((v-1)~\mathtt{mod}~m)+w \\
          s = \lceil \frac{v}{m} \rceil
  \end{array} \right.
  \label{eq:eq-5-1-1}
  \end{eqnarray}
\end{figure}

Relational-Sequential-Yu is in effect identical to the data
placement proposed by Yu et al.\,\cite{Yu07} in Section\,3.2.1.
Equation~(\ref{eq:eq-5-1-1}) is identical to the composition of
Equation~(\ref{eq:eq-3-5}) and Equation~(\ref{eq:eq-5-1-2}), i.e.,
$f_{MEMStoRS}(f_{RelationtoMEMS}(a_{v,w}))$. Thus, both
Relational-Sequential-Yu and Yu et al.'s data placement store the
attribute value $a_{v,w}$ in the same tip sector in the MEMS storage
device. Nevertheless, devising and understanding
Relational-Sequential-Yu is easier than coming up with Yu et al.'s
data placement since the RS model provides an abstraction of the
MEMS storage device.

\vspace*{-0.1cm}
\subsubsection{Relational-Parallel}
\vspace*{-0.1cm}

Relational-Parallel intends to provide highly parallel reading of
data by increasing the number of probe tips used during query
processing. Here, it is preferable that the values of the projected
attributes are placed as widely as possible in the direction of the
Region axis. Accordingly, Relational-Parallel stores the values of
each attribute such that a simultaneous-access sector group is
occupied with the values of only one attribute.

\begin{figure}[h!]
  \vspace*{0.50cm}
  \centerline{\psfig{file=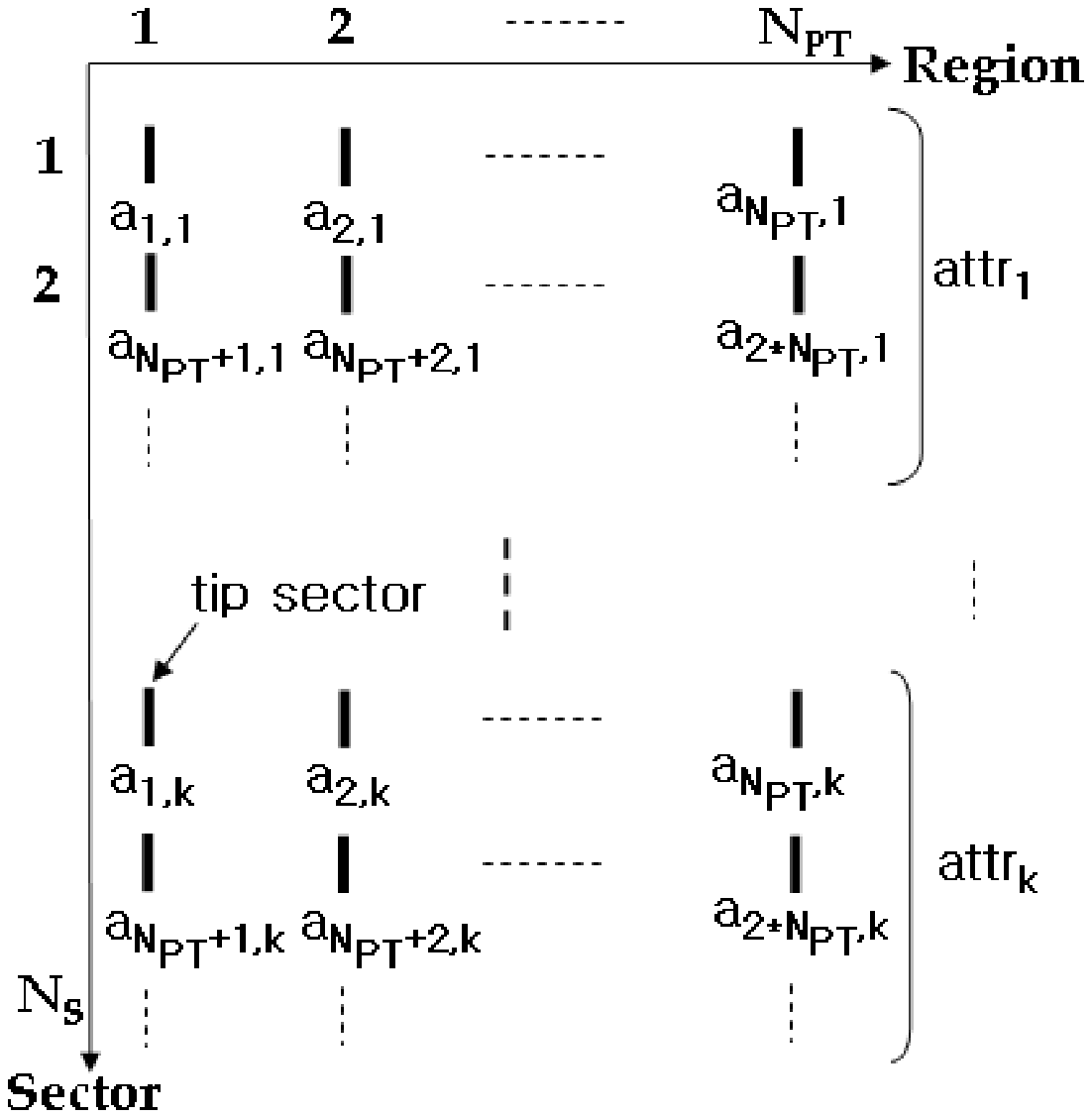, height=5.5cm}
              \hspace*{2.0cm}
              \psfig{file=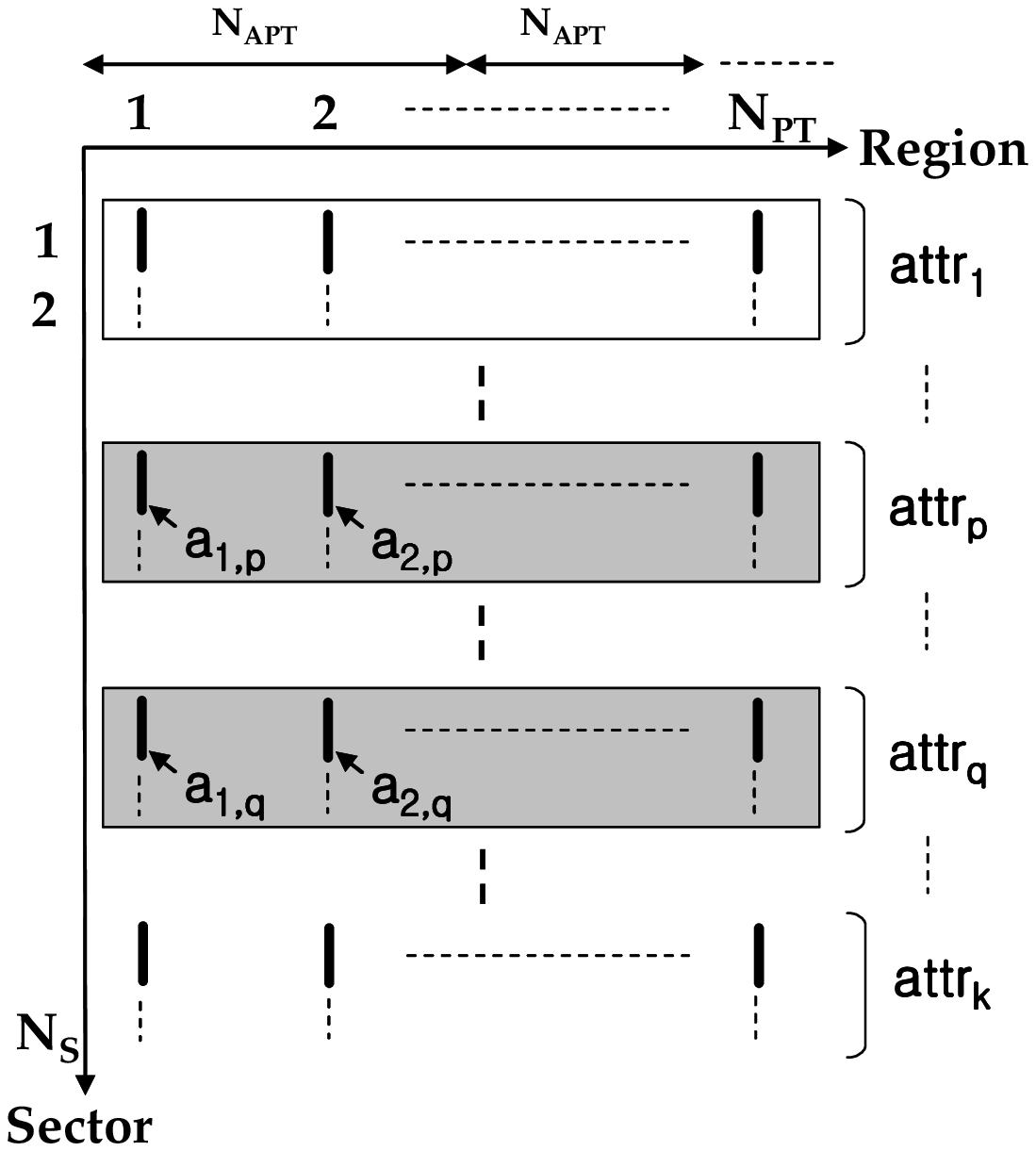, height=5.5cm}}
  \centerline{\hspace*{1.0cm} (a) Relational-Parallel. \hspace*{2.0cm} (b) The data area being retrieved by the query.}
  \vspace*{-0.2cm}
  \caption{Relational-Parallel data placement and the data area being retrieved by the query projecting $attr_{p}$ and $attr_{q}$.}
  \label{fig:5-RelationalTable-new}
\end{figure}

Figure~\ref{fig:5-RelationalTable-new} shows Relational-Parallel and
the data area being retrieved by the query. As shown in
Figure~\ref{fig:5-RelationalTable-new}(a), Relational-Parallel
stores the values of an attribute $attr_{p}$\, in a number of
successive simultaneous-access sector groups\,($1 \leq p \leq k$).
By such a placement, at most one seek operation occurs when reading
all the values of each attribute. In
Figure~\ref{fig:5-RelationalTable-new}(b), the shaded area indicates
the tip sectors accessed by the query projecting $attr_{p}$ and
$attr_{q}$. Since the width of the shaded area is $N_{PT}$, $\lceil
\frac{N_{PT}}{N_{APT}} \rceil$ sequential scans are required for
each attribute\,\footnote{\vspace*{-0.2cm} As in Footnote\,1, for
each scan, a turnaround operation occurs in practice, but it is
negligible compared with seek time or transfer time.}.

In order to show the excellence of Relational-Parallel, we deal with
another application that executes the range selection query in
Equation~(\ref{eq:eq-6-1}). This was also dealt with by Yu et
al.\,\cite{Yu07}

\begin{eqnarray}
& \textrm{SELECT} & attr_{1},~~attr_{p},~~attr_{q},~~... \nonumber \\
\label{eq:eq-6-1}
& \textrm{FROM}   & R \\
& \textrm{WHERE}  & attr_{1}~~>~~Bound; \nonumber
\end{eqnarray}

\begin{figure}[h!]
  \vspace*{0.50cm}
  \centerline{\psfig{file=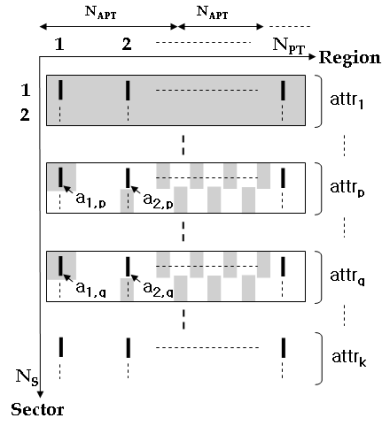, height=5.5cm}}
  \caption{The data area being retrieved by the range query projecting $attr_{1}$, $attr_{p}$, and $attr_{q}$.}
  \label{fig:5-RelationalTable-new2}
\end{figure}

\noindent Figure~\ref{fig:5-RelationalTable-new2} shows the data
area being retrieved by the range query. Relational-Parallel reads
the values of attributes as follows: (1) for the attribute in the
WHERE clause\,($attr_{1}$), it reads the value of every tuple, and
then, checks whether each tuple satisfies the condition $attr_{1} >
Bound$; (2) for the remaining attributes in a SELECT
clause\,($attr_{p}$, $attr_{q}$, ..., excluding $attr_{1}$), it
reads only those values that belong to the tuples satisfying the
condition. In Figure~\ref{fig:5-RelationalTable-new2}, the shaded
area indicates the tip sectors accessed by the range query
projecting $attr_{1}$, $attr_{p}$, and $attr_{q}$. $\lceil
\frac{N_{PT}}{N_{APT}} \rceil$ sequential scans are required for the
attribute $attr_{1}$; but only $\lceil
\frac{N_{PT}\,\times\,query\,selectivity}{N_{APT}} \rceil$ scans are
required for the attributes $attr_{p}$ and $attr_{q}$.

If relation $R$ has variable size attributes, both
Relational-Sequential-Yu and Relational-Parallel consider a variable
size attribute as a fixed size attribute with its maximum size as
was done by Yu et al.\,\cite{Yu07}.

Relational-Parallel is a new data placement that focuses on
parallelism, which is an important characteristic of the MEMS
storage device, while Relational-Sequential-Yu is the one that
focuses on reducing the number of seek operations.

\vspace*{-0.1cm}
\subsubsection{Comparison between Relation-Sequential-Yu and Relational-Parallel}
\vspace*{-0.1cm}

In data placements for relational data, the parameters affecting the
retrieval time are 1)\,the data size to be retrieved and 2)\,the
number of attributes to be projected. In this section, we compare
the retrieval time of Relational-Sequential-Yu and
Relational-Parallel by using Equation~(\ref{eq:eq-4-1}).
Table~\ref{tbl:analysis1} summarizes the notation used for analyzing
the retrieval time.

\vspace*{0.50cm}
\renewcommand{\baselinestretch}{1.10}
\begin{table}
\begin{center}
\caption{The notation used for analyzing the retrieval time.}
\vspace*{0.3cm}
\begin{tabular} {|c|l|c|}
\hline
\multicolumn{1}{|c|}{Symbols} & \multicolumn{1}{c|}{Definitions} \\
\hline \hline
$RetrievalDataSize$ & the data size to be retrieved for query processing (bytes) \\
\hline
$N_{projection}$ & the number of attributes to be projected by a query \\
\hline
$m$ & the number of tuples stored in one simultaneous-access sector group \\
    & in Relational-Sequential-Yu \\
\hline
\end{tabular}
\label{tbl:analysis1}
\end{center}
\end{table}
\renewcommand{\baselinestretch}{2.0}

For $TotalSeekTime$, Relational-Sequential-Yu is better than
Relational-Parallel. In Relational-Parallel, $K_{random} \leq
N_{projection}$ because at most $N_{projection}$ seek operations
could occur during query processing.
However, in Relational-Sequential-Yu, $K_{random} = 1$.


For $TotalTransferTime$, Relational-Parallel is better than
Relational-Sequential-Yu. In Relational-Sequential-Yu, $K_{parallel}
= min\,(m \times N_{projection},~N_{APT})$. On the other hand, in
Relational-Parallel, since $N_{PT}$ is usually a multiple of
$N_{APT}$\,\cite{Gri00-OSDI}, all $N_{APT}$ probe tips are used for
reading the data. Thus, $K_{parallel} = N_{APT}$.

The difference in $TotalTransferTime$ between the two data
placements increases as $RetrievalData\-Size$ gets lager, while the
difference in $TotlaSeekTime$ is limited to ($SeekTime_{rs} \times
N_{projection}$). Thus, as $RetrievalDataSize$ exceeds a certain
threshold, $RetrievalTime$ of Relational-Parallel becomes smaller
than that of Relational-Sequential-Yu because the advantage in the
transfer time overrides the disadvantage in the seek time.

\vspace*{-0.1cm}
\subsubsection{Comparison with Disk-Based Data Placements}
\vspace*{-0.1cm}

\begin{figure}[h!]
  \vspace*{0.50cm}
  \centerline{\psfig{file=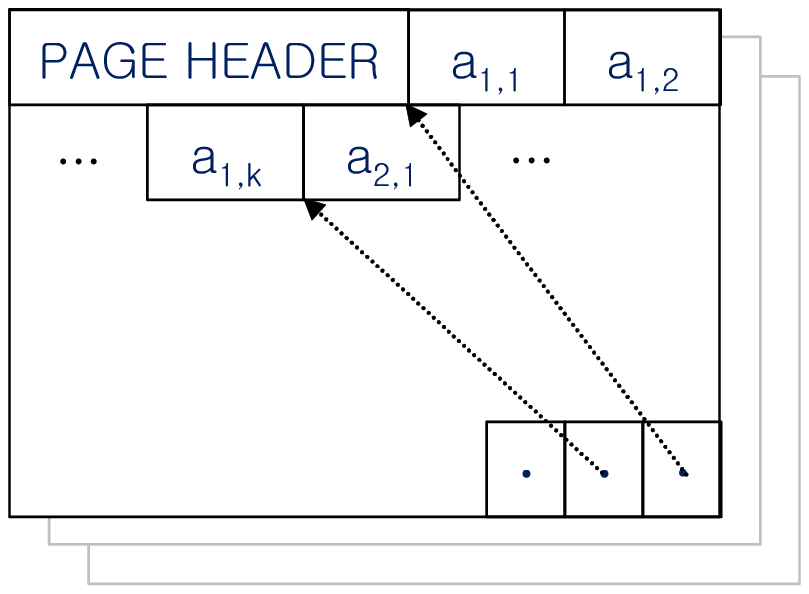, height=8cm}
              \hspace*{2.0cm}
              \psfig{file=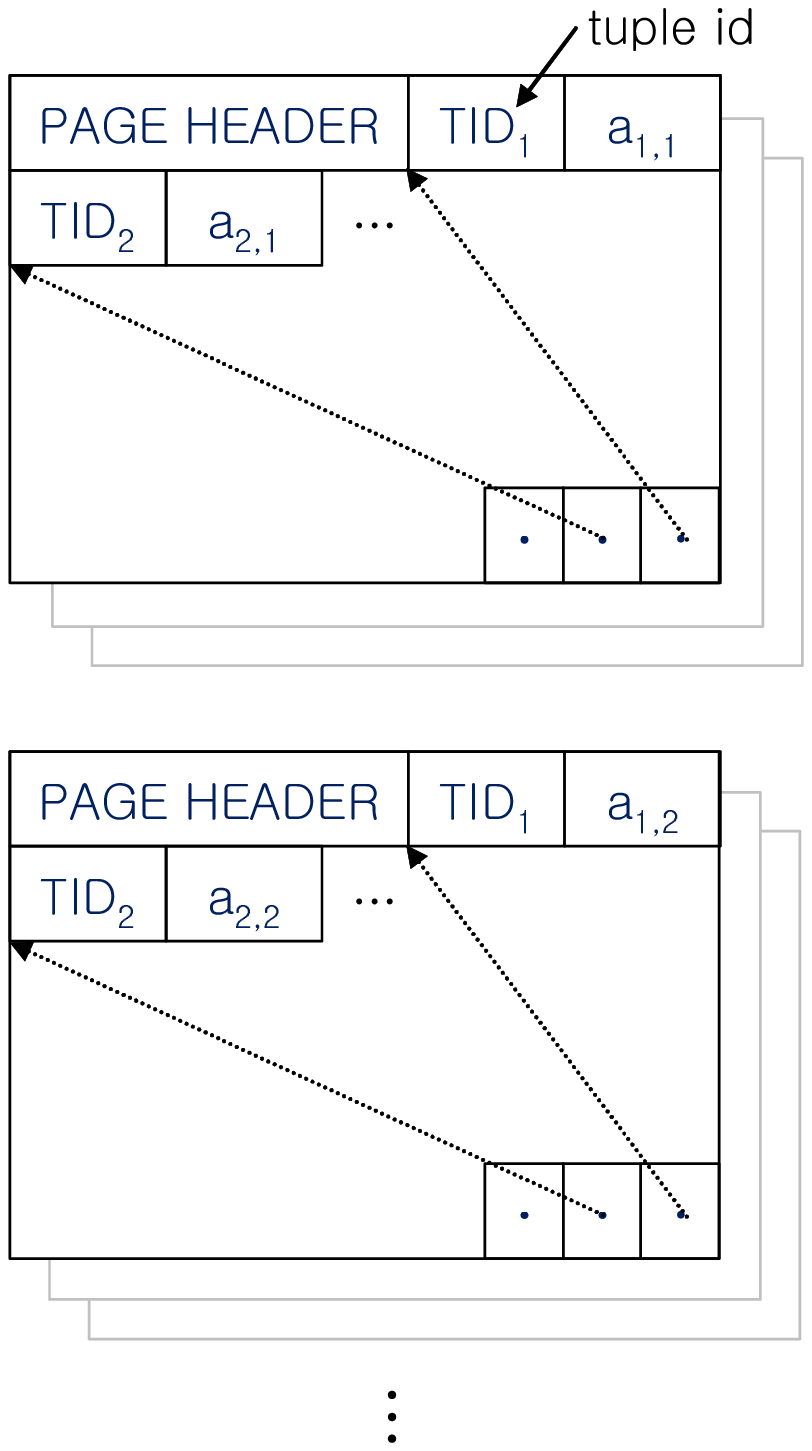, height=8cm}}
  \centerline{(a) NSM. \hspace*{5.0cm} (b) DSM.}
  \vspace*{-0.2cm}
  \caption{Data placements of the relation $R$ in slotted disk pages.}
  \label{fig:5-RelationalTable-Disk}
\end{figure}

Relational-Sequential-Yu and Relational-Parallel are similar to the
N-ary Storage Model\,(NSM)\,\cite{Ram00} and the Decomposition
Storage Model(DSM)\,\cite{Cop85}, respectively, which have been
proposed as data placements for relational data in a disk
environment. Figure~\ref{fig:5-RelationalTable-Disk} shows the data
placements of the relational $R$ by NSM and DSM. In
Figure~\ref{fig:5-RelationalTable-Disk}(a), NSM sequentially places
tuples of the relation $R$ in slotted disk pages. In
Figure~\ref{fig:5-RelationalTable-Disk}(b), DSM partitions a
relation $R$ into sub-relations based on the number of attributes
such that each sub-relation corresponds to an attribute. Here, DSM
places an attribute value of a tuple together with the identifier of
the tuple\,(simply, {\it TID}) so as to be used for joining
sub-relations.

Although the data placements of NSM and DSM are similar to those of
Relational-Sequential-Yu and Relational-Parallel, the data retrieval
costs for range select queries are quite different. As mentioned in
Section\,3, NSM and DSM consider $N_{APT}$ probe tips as one head.
But, Relational-Sequential-Yu and Relational-Parallel use multiple
probe tips for accessing data by freely selecting and activating
them. NSM reads all attribute values of the
tuples\,\cite{Ram00,Yu07}, while Relational-Sequential-Yu reads only
the projected attribute values by using multiple probe tips. DSM
reads all the values of the sub-relations corresponding to the
projected attributes\,\cite{Cop85,Yu07}, while Relational-Parallel
reads only those values of the tuples that satisfy the condition by
using multiple probe tips. However, if we consider the simple
projection queries with no range condition, Relational-Parallel
reads all the values of projected attributes as well. In this case,
Relational-Parallel becomes the same as DSM.

\vspace*{-0.1cm}
\subsection{Data Placements for Two-Dimensional Spatial Data}
\vspace*{-0.1cm}

In this section, we deal with an application that places a set of
two-dimensional spatial objects, and then, executes region queries
over those objects. This application is the same one dealt with by
Yu et al.\,\cite{Yu06} as described in Section\,3.2.2. We consider
two data placements for spatial data. We define the data placement
derived by using Strategy\_Sequential as {\it
Spatial-Sequential-Yu}, and the one derived by using
Strategy\_Parallel as {Spatial-Parallel}. Spatial-Sequential-Yu
turns out to be identical to the placement proposed by Yu et
al.\,\cite{Yu06}.

\vspace*{-0.1cm}
\subsubsection{Spatial-Sequential-Yu}
\vspace*{-0.1cm}

Spatial-Sequential-Yu intends to provide highly sequential reading
of data by preventing seek operations. We place spatial objects such
that a rectangular region in the two-dimensional space is
represented as a rectangular region in the RS model. By such a
placement, for any rectangular query region, we make $K_{random} =
0$ because objects in the query region are already
quasi-contiguously placed in the Sector axis of the RS
model\,\footnote{\vspace*{-0.2cm} If the number of objects along the
X axis exceeds $N_{APT}$ for the query region, more than one scan is
required. As in Footnote\,1, for each scan, a turnaround operation
occurs in practice, \vspace*{-0.2cm} but it is negligible compared
with seek time or transfer time.}.

\begin{figure}[h!]
  \vspace*{-0.50cm} 
  \vspace*{0.50cm}
  \centerline{\psfig{file=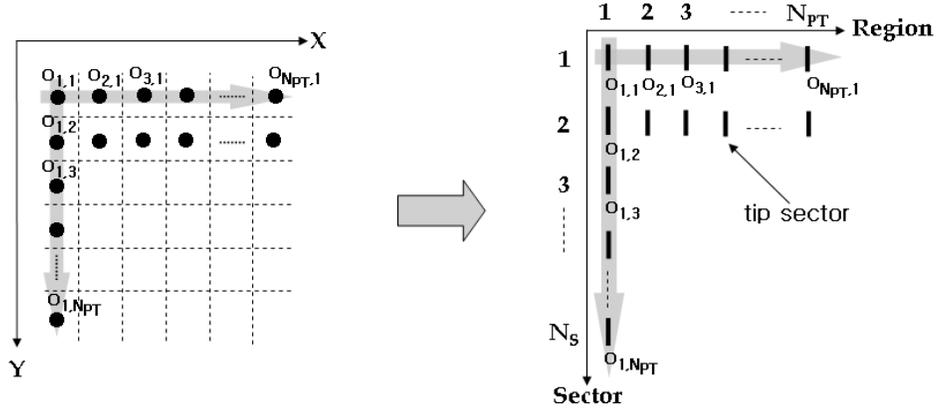, height=5.5cm}}
  \centerline{\hspace*{1.0cm} (a) The set $S$ of two-dimensional spatial objects. \hspace*{1.0cm} (b) Placement in the RS model.  \hspace*{2.5cm}}
  \vspace*{-0.2cm}
  \caption{Spatial-Sequential-Yu.}
  \label{fig:5-SpatialData-Yu}
\end{figure}

Figure~\ref{fig:5-SpatialData-Yu} shows Spatial-Sequential-Yu.
Spatial-Sequential-Yu places a spatial object in the X-Y plane on a
tip sector in the Region-Sector plane. Here, we again assume that
one spatial object can be stored in one tip sector.
Equation~(\ref{eq:eq-5-2-1}) shows a mapping function
$f_{SpacetoRS}$ that stores the object $o_{x,y}$ on the tip sector
$<$$r, s$$>$ in the RS model.

\clearpage 

\begin{figure}[h!]
  \begin{eqnarray}
  f_{SpacetoRS}\,(o_{x,y}) = \left\{ \begin{array}{l}
          r = x \\
          s = y
  \end{array} \right.
  \label{eq:eq-5-2-1}
  \end{eqnarray}
\end{figure}

\begin{figure}[h!]
  \vspace*{0.50cm}
  \centerline{\psfig{file=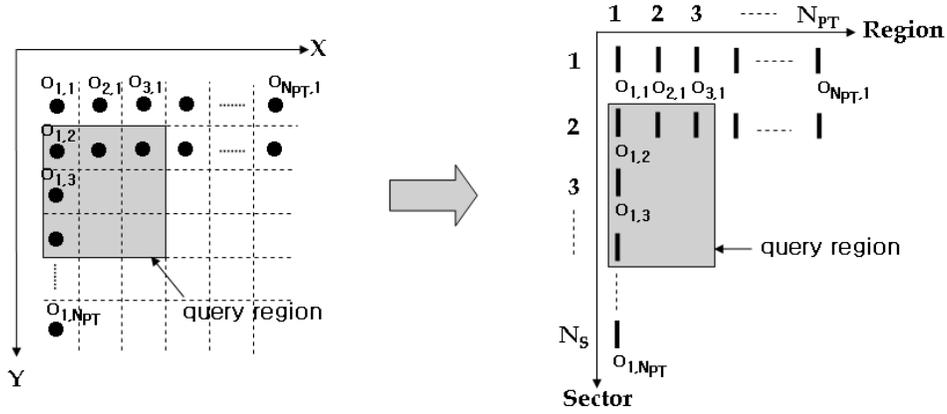, height=5.5cm}}
  \centerline{(a) The query region to be retrieved \hspace*{2.5cm} (b) The query region to be retrieved}
  \vspace*{-0.2cm}\centerline{in the two-dimensional space. \hspace*{4.0cm} ~~~in the RS model.~~~~~}
  \vspace*{-0.2cm}
  \caption{The query region to be retrieved in Spatial-Sequential-Yu.}
  \label{fig:5-SpatialData-Yu2}
\end{figure}

In Figure~\ref{fig:5-SpatialData-Yu2}(a), the shaded area indicates
the query region in the two-dimensional space. In
Figure~\ref{fig:5-SpatialData-Yu2}(b), the shaded area indicates the
corresponding region in the RS model. Let $QueryRegionSize_{x}$ be
the width of the corresponding query region. Then, $\lceil
\frac{QueryRegionSize_{x}}{N_{APT}} \rceil$ sequential scans are
required for query processing.

If the number of spatial objects in the direction of the X axis is
larger than $N_{PT}$, we vertically partition the two-dimensional
space into components having a width of $N_{PT}$ or less, and then,
place the components on the Region-Sector plane along the direction
of the Sector axis. Then, the query cost should reflect one
additional seek time for each component.

Spatial-Sequential-Yu is in effect identical to the data placement
proposed by Yu et al.\,\cite{Yu06} in Section\,3.2.2.
Equation~(\ref{eq:eq-5-2-1}) is identical to the composition of
Equation~(\ref{eq:eq-3-5}) and Equation~(\ref{eq:eq-5-2-2}), i.e.,
$f_{MEMStoRS}(f_{SpacetoMEMS}(o_{x,y}))$. Thus, both
Spatial-Sequential-Yu and Yu et al.'s data placement put the object
$o_{x,y}$ in the same tip sector in the MEMS storage device.
Nevertheless, as in Relational-Sequential-Yu, understanding
Spatial-Sequential-Yu is much easier than understanding Yu et al.'s
data placement due to the abstraction of the RS model.

\vspace*{-0.1cm}
\subsubsection{Spatial-Parallel}
\vspace*{-0.1cm}

Spatial-Parallel intends to provide highly parallel reading of data
by increasing the number of probe tips used during query processing.
We partition the two-dimensional space into blocks, and then, place
spatial objects in a block into a simultaneous-access sector group
of the RS model. By such a placement, for any rectangular query
region, we can make $K_{parallel}$ to be as close to $N_{APT}$ as
possible.

\begin{figure}[h]
  \vspace*{0.50cm}
  \centerline{\psfig{file=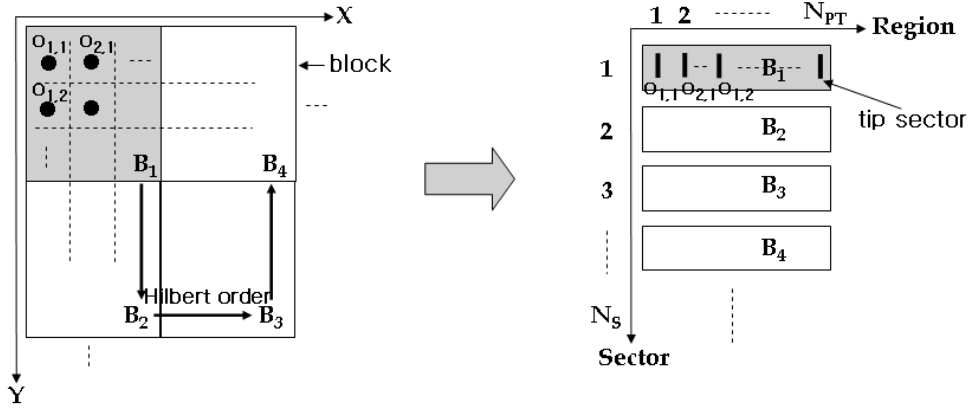, height=5.5cm}}
  \centerline{(a) The set $S$ of two-dimensional spatial objects. \hspace*{1.0cm} (b) Placement in the RS model. \hspace*{1.0cm}}
  \vspace*{-0.2cm}
  \caption{Spatial-Parallel.}
  \label{fig:5-SpatialData-parallel}
  \vspace*{0.5cm} 
\end{figure}

Figure~\ref{fig:5-SpatialData-parallel} shows Spatial-Parallel,
which places spatial objects through the following three steps.

\vspace*{-0.3cm}
\begin{description}
\item [1. Partitioning step:]
We partition the two-dimensional space into blocks that form a
rectangular grid such that the total size of spatial objects in one
block is equal to the total size of tip sectors in one simultaneous
sector group.

\item [2. Ordering step:]
We sort the partitioned blocks according to a space filling
curve\,\cite{Jag90}. A space filling curve such as the
Z-order\,\cite{Ore86} or Hilbert order\,\cite{Hil91,Moo01}, is a way
of linearly ordering regions in a multi-dimensional space into a
one-dimensional space so as to keep the clustering\,\cite{Jag90}.
Here, We use the Hilbert order.

\item [3. Placement step:]
We place spatial objects of the $i$\,th block in the sequence
constructed in Step 2 on the $i$\,th simultaneous-access sector
group of the RS model in the row-major order\,($1 \leq i \leq
N_{block}$).
\end{description}

\begin{figure}[h]
  \vspace*{-0.5cm} 
  \vspace*{0.50cm}
  \centerline{\psfig{file=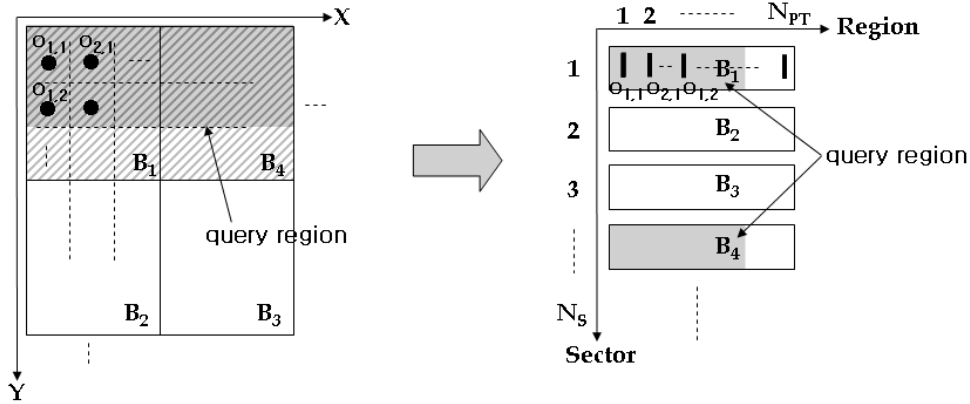, height=5.5cm}}
  \centerline{(a) A two-dimensional space. \hspace*{2.5cm} (b) The RS model.}
  \vspace*{-0.2cm}
  \caption{The query region to be retrieved in Spatial-Parallel.}
  \label{fig:5-SpatialData-parallel2}
\end{figure}

Figure~\ref{fig:5-SpatialData-parallel2} shows the region being
retrieved by a query. In
Figure~\ref{fig:5-SpatialData-parallel2}(a), the shaded area
indicates the query region, and the slashed area indicates the set
of blocks overlapping with the query region. Hereafter, we call this
set of overlapping blocks the $QueryBlockSet$. In
Figure~\ref{fig:5-SpatialData-parallel2}(b), the shaded area
indicates the corresponding query region to be retrieved in the RS
model. For data retrieval, we first find the set of
simultaneous-access sector groups corresponding to $QueryBlockSet$,
and then, read the data on tip sectors overlapping with the query
region\,\footnote{\vspace*{-0.2cm} If the number of tip sectors
overlapping with the query region exceeds $N_{APT}$, more than one
scan is required. As in Footnote\,1, for each scan, a turnaround
operation occurs in practice, \vspace*{-0.2cm} but it is negligible
compared with seek time or transfer time.}. Here, seek operations
occur at most as many times as the number of blocks in the
$QueryBlockSet$.

Here, we use two physical database design techniques to reduce the
number of seek operations during query processing. First, in the
partitioning step, we set the aspect ratio of a
block\,($BlockAspectRatio$) to be the weighted average aspect ratio
of a query region defined as $QueryAspectRatio =
\frac{\sum_{i}\,(f_{i} \times
QueryResionSize_{i_{x}})}{\sum_{i}\,(f_{i} \times
QueryResionSize_{i_{y}})}$, where $f_{i}$ is the query frequency. It
has been proven by Lee et al.\,\cite{Lee97} that the number of
blocks in $QueryBlockSet$ is minimized when this condition is met.
Second, in the ordering step, we use the Hilbert order as the space
filling curve. The more contiguously the simultaneous-access sector
groups corresponding to $QueryBlockSet$ are placed, the fewer seek
operations occur during query processing. Here, the degree of
clustering of the blocks in $QueryBlockSet$ is dependent on the
space filling curve to be used. It is known that the Hilbert order
achieves the best clustering\,\cite{Moo01}.

Spatial-Parallel is a new data placement technique that focuses on
parallelism, while Spatial-Sequential-Yu focuses on reducing the
number of seek operations as in the traditional disk-based approach.

\vspace*{-0.1cm}
\subsubsection{Comparison between Spatial-Sequential-Yu and Spatial-Parallel}
\vspace*{-0.1cm}

The parameters affecting the retrieval time in data placements for
two-dimensional spatial data are the size and the aspect ratio of
the query region. In this section, we compare the retrieval time of
Spatial-Sequential-Yu and Spatial-Parallel by using
Equation~(\ref{eq:eq-4-1}). Table~\ref{tbl:analysis2} summarizes the
notation to be used for analyzing the retrieval time.

\vspace*{0.50cm}
\renewcommand{\baselinestretch}{1.10}
\begin{table}
\begin{center}
\caption{The notation for analyzing the retrieval time.}
\vspace*{0.3cm}
\begin{tabular} {|c|l|c|}
\hline
\multicolumn{1}{|c|}{Symbols} & \multicolumn{1}{c|}{Definitions} \\
\hline \hline
$QueryRegionSize_{x}$ & the width of a query region \\
\hline
$QueryRegionSize_{y}$ & the height of a query region \\
\hline
$QueryRegionSize$ & the size of a query region (= $QueryRegionSize_{x} \times QueryRegionSize_{y}$) \\
\hline
$QueryAspectRatio$ & the ratio of width to height of a query region (= $\frac{QueryRegionSize_{x}}{QueryRegionSize_{y}}$) \\
\hline
$\#QueryBlocks$ & the number of blocks in $QueryBlockSet$ \\
\hline
\end{tabular}
\label{tbl:analysis2}
\end{center}
\end{table}
\renewcommand{\baselinestretch}{2.0}

For $TotalSeekTime$, Spatial-Sequential-Yu is better than
Spatial-Parallel. For Spatial-Sequential-Yu, $K_{random} = 1$
because a query region is retrieved without seek operations. For
Spatial-Parallel, $K_{random} \leq \#QueryBlocks$. Thus, from
Equation~(\ref{eq:eq-4-1}), Spatial-Parallel has additional seek
time of at most $\#QueryBlocks \times SeekTime_{rs}$ compared with
Spatial-Sequential-Yu.

For $TotalTransferTime$, either Spatial-Sequential-Yu or
Spatial-Parallel is better than the other depending on the size and
aspect ratio of the query region. In Spatial-Sequential-Yu,
$K_{parallel}$ decreases as $QueryRegionSize$ or $QueryAspectRatio$
gets smaller because less probe tips can be used to read the tip
sectors in the query region. On the other hand, in Spatial-Parallel,
$K_{parallel}$ is less affected by $QueryAspectRatio$ than in
Spatial-Sequential-Yu because a query region is represented as a set
of simultaneous-access sector groups rather than as a rectangular
region. For example, when $QueryAspectRatio$ is very small\,(e.g.,
$QueryAspectRatio = \frac{1}{16}$), in Spatial-Sequential-Yu, only a
few probe tips may be used; but in Spatial-Parallel, much more probe
tips will be used because objects in the query region are placed
widely in the direction of the Region axis. Therefore,
Spatial-Parallel has more advantage over Spatial-Sequential-Yu as
$QueryRegionSize$ or $QueryAspectRatio$ gets smaller.

If $QueryRegionSize$ or $QueryAspectRatio$ decreases below a certain
threshold, the retrieval time of Spatial-Parallel becomes smaller
than that of Spatial-Sequential-Yu because its advantage in the
transfer time more than compensates for its disadvantage in seek
time. Consequently, Spatial-Parallel has the following two good
characteristics: (1)\,the data retrieval performance is superior to
that of Spatial-Sequential-Yu for highly selective queries, (2)\,the
performance is largely independent of the aspect ratio of the query
region.

%
%
\section{Performance Evaluation}
\vspace*{-0.30cm}

\vspace*{-0.1cm}
\subsection{Experimental Data and Environment}
\vspace*{-0.1cm}

We compare the data retrieval performance of the new data placements
proposed in this paper with those of existing data placements. We
use retrieval time as the measure of the performance.

\vspace*{-0.1cm}
\subsubsection{Experiments for Relational Data}
\vspace*{-0.1cm}

We compare data retrieval performance of the following five data
placements: Relational-Parallel, Relational-Sequential-Yu,
Relational-LowerBound, NSM-Griffin, and DSM-Griffin. Here, {\it
Relational-LowerBound} is a virtual data placement that has a lower
bound of retrieval time in the RS model\,(i.e., $K_{parallel} =
N_{APT}$ and $K_{random} = 0$). We use this data placement in order
to show how close the performance of each of the other data
placements is to a lower bound of the RS model. {\it NSM-Griffin}
and {\it DSM-Griffin} are the data placements using
NSM\,\cite{Ram00} and DSM\,\cite{Cop85} in Section\,6.1.4 based on
the linear abstraction proposed by Griffin et
al.\,\cite{Gri00-OSDI}, which corresponds to the disk mapping layer
of Figure~\ref{fig:3-SystemArchitecture}(a). In NSM-Griffin and
DSM-Griffin, $N_{APT}$ probe tips are activated for accessing data.


For experimental data, we use the synthetic relational data that is
used by Yu et al.\,\cite{Yu07}. Here, we set the number of
attributes of the relation to be 16 and the size of each attribute
to be 8\,bytes as in Yu et al.\,\cite{Yu07}.

We perform two experiments for the range selection query in
Equation~(\ref{eq:eq-6-1}). In Experiment 1, we measure the
retrieval time while varying data size from 5\,Mbytes to
320\,Mbytes. Here, we set $N_{projection} = 8$ and selectivity =
0.1.
In Experiment 2, we measure the retrieval time while varying
$N_{projection}$ from $1$ to $16$.
Table~\ref{tbl:parameter-relation} summarizes these experiments and
the parameters.

\vspace*{0.50cm}
\renewcommand{\baselinestretch}{1.10}
\begin{table}
\begin{center}
\caption{Experiments and parameters for relational data.}
\vspace*{0.3cm}
\begin{tabular} {|c|c|c|c|}
\hline
\multicolumn{2}{|c|}{Experiments} & \multicolumn{2}{c|}{Parameters} \\
\hline \hline Experiment 1& comparison of data retrieval performance
& data size        & 5 $\sim$ 320\,Mbytes \\ \cline{3-4}
            & as the size of data is varied            & $N_{projection}$ & 8 \\
\hline Experiment 2& comparison of data retrieval performance & data
size        & 320\,Mbytes        \\ \cline{3-4}
            & as $N_{projection}$ is varied            & $N_{projection}$ & 1 $\sim$ 16          \\
\hline
\end{tabular}
\label{tbl:parameter-relation}
\end{center}
\end{table}
\renewcommand{\baselinestretch}{2.0}

\vspace*{-0.1cm}
\subsubsection{Experiments for Two-Dimensional Spatial Data}
\vspace*{-0.1cm}

Here, we compare data retrieval performance of three data
placements: Spatial-Parallel, Spatial-Sequential-Yu, and
Spatial-LowerBound. As in Section\,7.1.1, {\it Spatial-LowerBound}
is defined to be the case where $K_{parallel} = N_{APT}$ and
$K_{random} = 0$.

For the experimental data, we use the synthetic spatial data that is
generated by the same method used by Yu et al.\,\cite{Yu06}. Here,
we set the number of spatial objects to be $40,960,000$ and the size
of each object to be 8\,bytes.

We perform two experiments. In Experiment 3, we measure the
retrieval time while varying $QueryRegionSize$ from $0.01\%$ to
$10\%$ of that of the spatial data. Here, the shape of a query is a
square\,(i.e., $QueryAspectRatio = 1$). In Experiment 4, we measure
the retrieval time while varying $QueryAspectRatio$ from $16$ to
$\frac{1}{16}$. Here, we fix $QueryRegionSize$ to be $1\%$ of the
size of the spatial data. Table~\ref{tbl:parameter-spatial}
summarizes the experiments and the parameters.

\vspace*{0.50cm}
\renewcommand{\baselinestretch}{1.10}
\begin{table}
\begin{center}
\caption{Experiments and parameters for two-dimensional spatial
data.} \vspace*{0.3cm}
\begin{tabular} {|c|c|c|c|}
\hline
\multicolumn{2}{|c|}{Experiments} & \multicolumn{2}{c|}{Parameters} \\
\hline \hline Experiment 3 & comparison of data retrieval
performance & $QueryRegionSize $ & $0.01 \sim 10\,\%$ \\ \cline{3-4}
             & as $QueryRegionSize$ is varied               & $QueryAspectRatio$ & 1 \\
\hline Experiment 4 & comparison of data retrieval performance &
$QueryRegionSize$  & 1\,\% \\ \cline{3-4}
             & as $QueryAspectRatio$ is varied              & $QueryAspectRatio$ & $16 \sim \frac{1}{16}$ \\
\hline
\end{tabular}
\label{tbl:parameter-spatial}
\end{center}
\end{table}
\renewcommand{\baselinestretch}{2.0}

\vspace*{-0.1cm}
\subsubsection{An Emulator of the MEMS Storage Device}
\vspace*{-0.1cm}

We have implemented an emulator of the MEMS storage device since a
physical MEMS storage device is not available on the market yet. We
have implemented an emulator of the CMU MEMS storage device using
formulas and parameters proposed by Griffin et
al.\,\cite{Gri00-OSDI,Gri00-SIGMETRICS}. We conduct all experiments
on a Pentium\,4 3.0\,GHz Linux PC with 2\,GBytes of main memory.

\vspace*{-0.1cm}
\subsection{Results of the Experiments}
\vspace*{-0.1cm}

\vspace*{-0.1cm}
\subsubsection{Relational Data}
\vspace*{-0.1cm}

\begin{figure}[h!]
  \vspace*{0.50cm}
  \centerline{\psfig{file=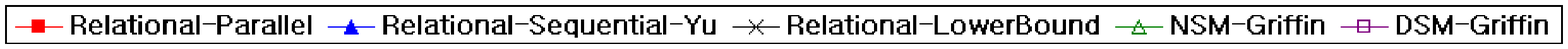, height=0.4cm}}
  \vspace*{0.3cm}
  \centerline{\psfig{file=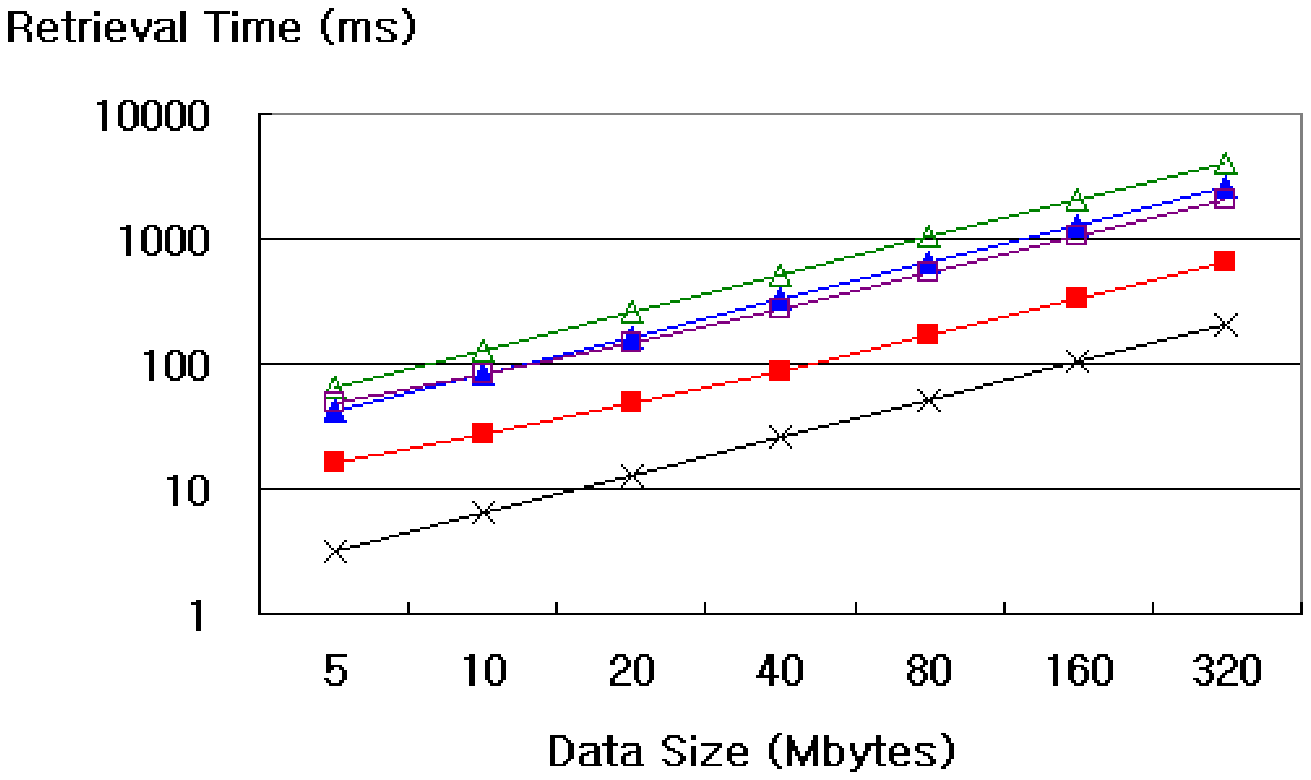, height=5cm}}
  \vspace*{-0.2cm}
  \caption{Retrieval time for relational data as the data size is varied\,($N_{projection} = 8$, selectivity = 0.1).}
  \label{fig:6-RelationalTable}
\end{figure}

Figure~\ref{fig:6-RelationalTable} shows the retrieval time of five
data placements as the data size is
varied\,\footnote{\vspace*{-0.2cm} Here, for the sake of fairness,
we did not include the TIDs in DSM-Griffin that are used for joins.
Our method Relational-Parallel and Relational-Sequential-Yu do not
use TIDs \vspace*{-0.2cm} since we use the maximum size for variable
size attributes.}. As analyzed in Section\,6.1, Relational-Parallel
is superior to Relational-Sequential-Yu.
As the size of data is varied from 5\,Mbytes to 320\,Mbytes, the
performance of Relational-Parallel improves from 2.6 to 4.0 times
over that of Relational-Sequential-Yu. We note that the query
performance of NSM-Griffin is much poorer than those of the others.
This result indicates that disk mapping approaches provide
relatively poor performance compared with device-specific approaches
since the characteristics of the MEMS storage device are not fully
utilized.

\begin{figure}[h!]
  \vspace*{0.50cm}
  \centerline{\psfig{file=6-RelationalTable-Result-caption3.eps, height=0.4cm}}
  \vspace*{0.3cm}
  \centerline{\psfig{file=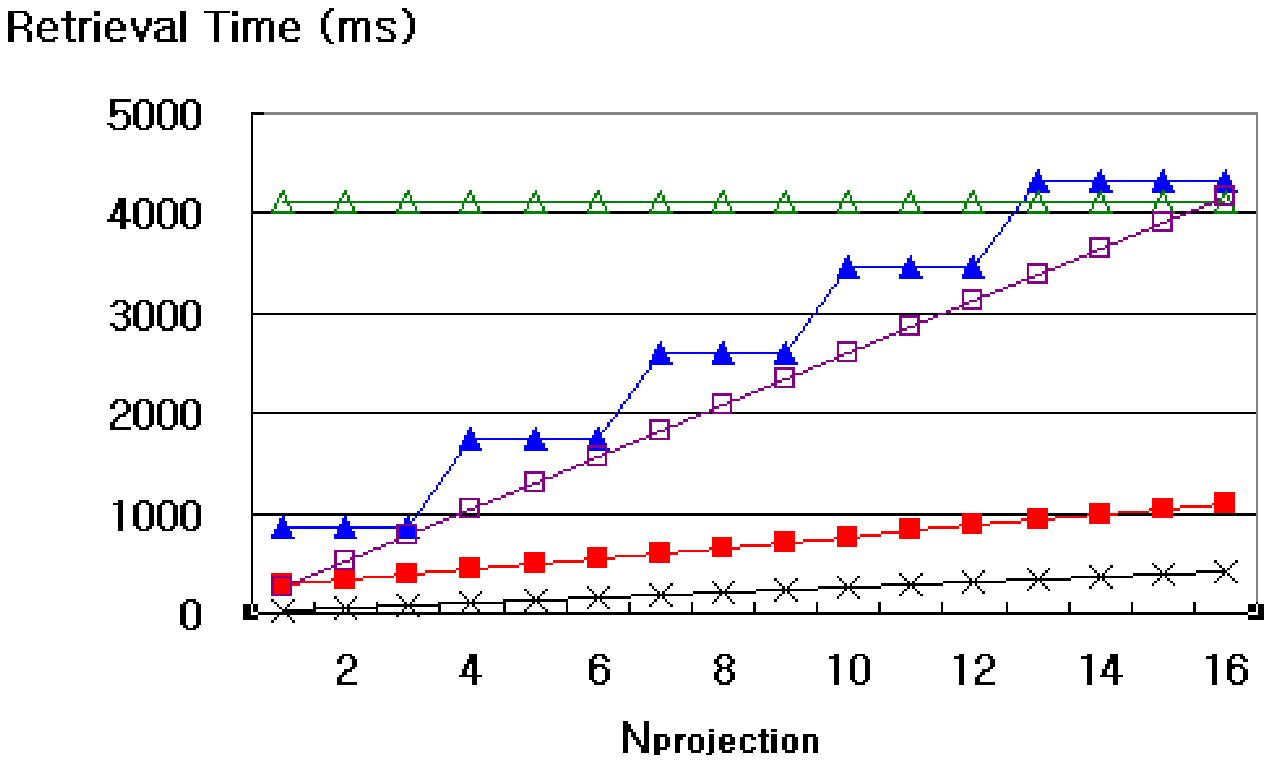, height=5cm}}
  \vspace*{-0.2cm}
  \caption{Retrieval time for relational data as $N_{projection}$ is varied\,(selectivity = 0.1).}
  \label{fig:6-RelationalTable2}
\end{figure}

Figure~\ref{fig:6-RelationalTable2} shows the retrieval time of five
data placements as $N_{projection}$ is varied. As $N_{projection}$
increases, the retrieval time of Relational-Parallel increases
linearly. In contrast, that of Relational-Sequential-Yu increases in
a stepwise manner. The reason for this behavior is that the number
of sequential scans\,($\lceil \frac{m \times
N_{projection}}{N_{APT}} \rceil$) in Relational-Sequential-Yu
increases by an integer number.
We note that Relational-Parallel is closer to Relational-LowerBound
than Relational-Sequential-Yu. The retrieval time of NSM-Griffin is
constant over all $N_{projection}$ because it always reads all the
attribute values of the relation regardless of $N_{projection}$.

In Figure~\ref{fig:6-RelationalTable2}, we note that the retrieval
time of Relational-Sequential-Yu is slightly larger than those of
NSM-Griffin and DSM-Griffin when accessing the entire
relation\,(i.e., $N_{projection} = 16$).
It is because the linear abstraction proposed by Griffin et
al.\,\cite{Gri00-OSDI} is optimized for sequential access. The
linear abstraction arranges tip sectors so as to fast access all the
tip sectors in the MEMS storage device. It first accesses all the
tip sectors of the first column of every region by activating
another set of $N_{APT}$ probe tips, and then, accesses all the tip
sectors of the second column, and so on. Thus, when accessing the
entire tip sectors in the MEMS storage device, the RS model is worse
than the linear abstraction in seek time. The number of seek
operations of the RS model\,($S_{x} \times \lceil
\frac{N_{PT}}{N_{APT}} \rceil$) is larger than that of the linear
abstraction\,($S_{x}$).



\vspace*{0.7cm} 
\vspace*{-0.1cm}
\subsubsection{Two-Dimensional Spatial Data}
\vspace*{-0.1cm}

\begin{figure}[h!]
  \vspace*{0.50cm}
  \centerline{\psfig{file=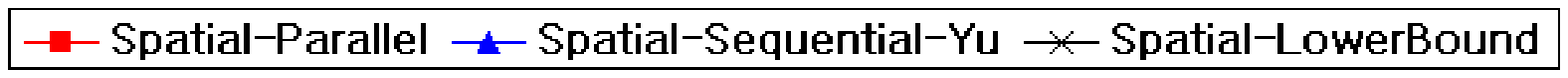, height=0.4cm}}
  \vspace*{0.3cm}
  \centerline{\psfig{file=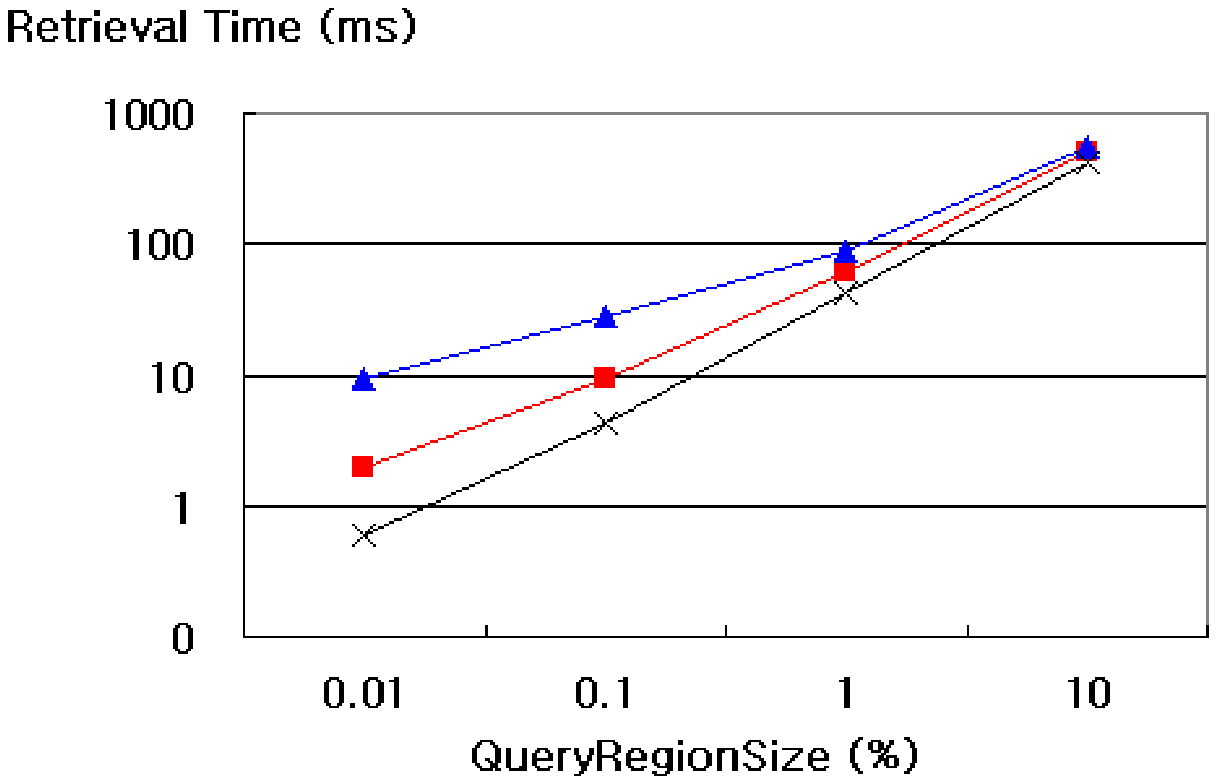, height=5cm}}
  \vspace*{-0.2cm}
  \caption{Retrieval time of spatial data as $QueryRegionSize$ is varied\,($QueryAspectRatio = 1$).}
  \label{fig:6-SpatialData}
\end{figure}

Figure~\ref{fig:6-SpatialData} shows the retrieval time of three
data placements as $QueryRegionSize$ is varied. As we argued in
Section\,6.2, we observe that Spatial-Parallel becomes superior to
Spatial-Sequential-Yu as $QueryRegionSize$ gets smaller, that is, as
the selectivity of the query gets lower. In
Figure~\ref{fig:6-SpatialData}, as $QueryRegionSize$ is varied from
$10\,\%$ to $0.01\,\%$, the performance of Spatial-Parallel improves
from $1.1$ to $4.8$ times over that of Spatial-Sequential-Yu.

\begin{figure}[h!]
  \vspace*{0.50cm}
  \centerline{\psfig{file=6-SpatialData-Result-caption.eps, height=0.4cm}}
  \vspace*{0.3cm}
  \centerline{\psfig{file=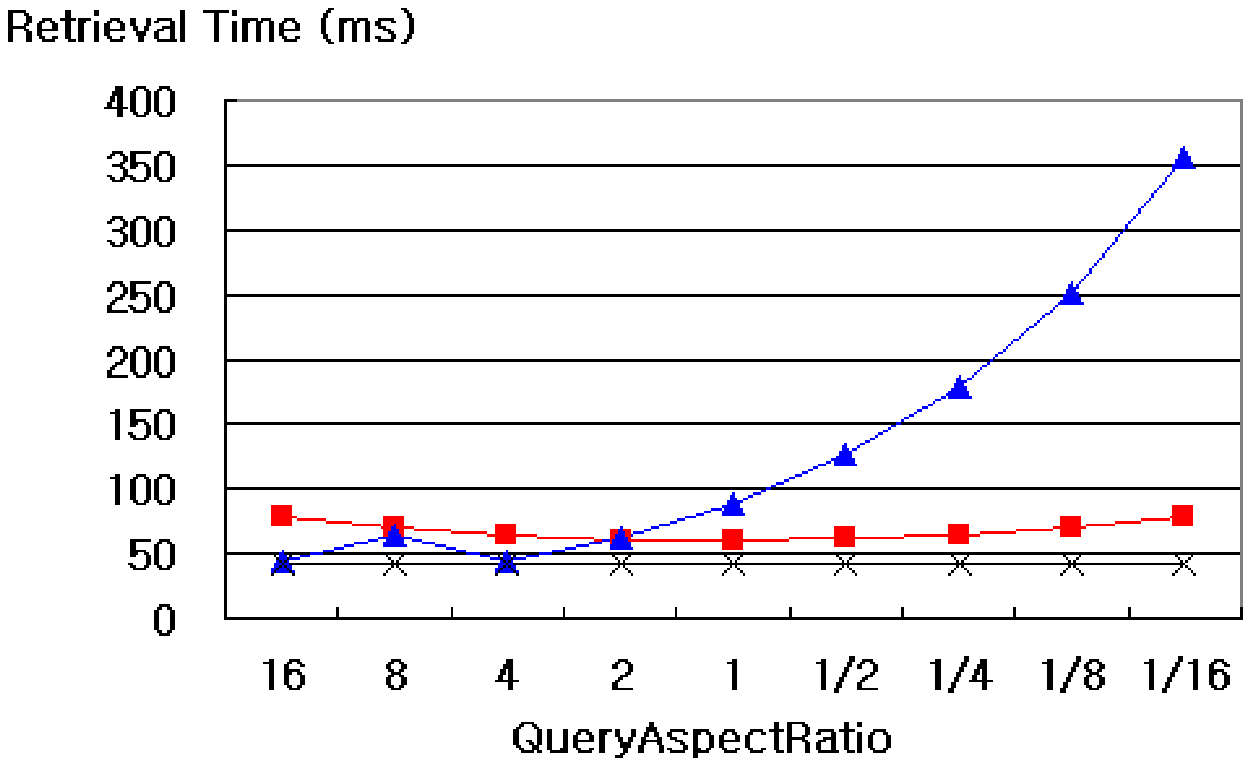, height=5cm}}
  \vspace*{-0.2cm}
  \caption{Retrieval time of spatial data as $QueryAspectRatio$ is varied\,($QueryRegionSize = 1\%$).}
  \label{fig:6-SpatialData2}

\end{figure}

Figure~\ref{fig:6-SpatialData2} shows the retrieval time as
$queryAspectRatio$ is varied. As we argued in Section\,6.2, we
observe that Spatial-Sequential-Yu degrades as $QueryAspectRatio$
decreases\,(i.e., $QueryRegionSize_{x}$ decreases). This is because
$K_{parallel}$ in Spatial-Sequential-Yu decreases. The performance
of Spatial-Parallel, however, stays largely flat regardless of
$QueryAspectRatio$. Figure~\ref{fig:6-SpatialData2} also shows that
Spatial-Parallel is close to Spatial-LowerBound.

In Figure~\ref{fig:6-SpatialData2}, we note that the retrieval time
of Spatial-Sequential-Yu when $QueryAspectRatio = 8$ is slightly
larger than the time when $QueryAspectRatio = 4$. It is because the
case of $QueryAspectRatio = 8$ requires more scan operations for
accessing the query region than that of $QueryAspectRatio = 4$. The
case of $QueryAspectRatio = 8$ requires two scans\,($\lceil
\frac{1820}{1280} \rceil = 2$) as mentioned in Section\,6.2 while
the case of $QueryAspectRatio = 4$ only one scan\,($\lceil
\frac{1280}{1280} \rceil = 1$). Although the case of
$QueryAspectRatio = 16$ also requires two scans\,($\lceil
\frac{2560}{1280} \rceil = 2$), it takes less retrieval time than
the case of $QueryAspectRatio = 8$ because the height of the query
region\,(i.e., $QueryRegionSize_{y}$) is shorter than the case of
$QueryAspect\-Ratio = 8$.

%
%
\section{Conclusions}
\vspace*{-0.30cm}

In this paper, we have proposed a logical model called the RS model
that abstracts the physical MEMS storage model. The RS model
simplifies the structure of the MEMS storage device by rearranging
its tip sectors into a virtual two-dimensional plane. As a result,
the RS model represents the position of a tip sector with only two
parameters while the physical MEMS storage model requires four
parameters. Despite this simplification, the RS model provides
characteristics for random access and sequential access\,(i.e., seek
time and transfer rate) almost identical to those of the physical
MEMS storage model.

We have presented an analytic formula for retrieval performance of
the RS model in Equation~(\ref{eq:eq-4-1}), and then, proposed
heuristic data placement strategies -- Strategy\_Sequential and
Strategy\_Parallel -- based on that formula. Strategy\_Parallel
intends to maximize the number of probe tips to be used while
Strategy\_Sequential intends to minimize the number of seek
operations.

By using those strategies, we have derived data placements for
relational data and two-dimensional spatial data. We have identified
that data placements derived by Strategy\_Sequential are in effect
identical to those in Yu et al.\,\cite{Yu06,Yu07} and that those
derived by Strategy\_Parallel are new ones discovered. Further,
through extensive analysis and experiments, we have compared the
retrieval performance of our new data placements with those of
existing ones. Experimental results using relational data of
320\,MBytes\, show that Relational-Parallel improves the performance
by up to 4.0 times\,(where $N_{projection} = 8$ and the query
selectivity = 0.1) compared with Yu et
al.\,\cite{Yu07}\,(Relational-Sequential-Yu). This performance gain
would be even higher for smaller query selectivities. Experimental
results using two-dimensional spatial data of 328\,MBytes\, also
show that Spatial-Parallel improves data retrieval performance by up
to 4.8 times\,(where $QueryRegionSize = 0.01\,\%$ and
$QueryAspectRatio = 1$) compared with Yu et
al.\,\cite{Yu06}\,(Spatial-Sequential-Yu). Furthermore, these
improvements are expected to become more marked as the size of the
data grows, reflecting the strength of our model.

Overall, these results indicate that the RS model is a new logical
model for the MEMS storage device that allows users to easily
understand and effectively use this rather complex device.

%
%
\section{Acknowledgement}
\vspace*{-0.3cm}




We would like to thank Dr. Young-Koo Lee of Kyung Hee University in
Korea for his helpful advice and discussions. This work was
supported by the Korea Science and Engineering Foundation\,(KOSEF)
grant funded by the Korea
government(MEST)\,(No.~R0A-2007-000-20101-0).

%
%

\end{document}